\documentclass[12pt,preprint]{aastex}
\usepackage{amssymb, amsmath, apjfonts, emulateapj5, natbib, color}
\input psfig.tex

\def\aa{{A\&A}}

\def\aj{{AJ}}

\def\apj{{ApJ}}
\def\apjs{{ApJS}}
\def\asec{$^{\prime\prime}$}

\def\hb{H$\beta$}
\def\hst{{\it HST}}
\def\kms{km s$^{-1}$}
\def\lamb{$\lambda$}
\def\lax{{$\mathrel{\hbox{\rlap{\hbox{\lower4pt\hbox{$\sim$}}}\hbox{$<$}}}$}}
\def\gax{{$\mathrel{\hbox{\rlap{\hbox{\lower4pt\hbox{$\sim$}}}\hbox{$>$}}}$}}
\def\simlt{\lower.5ex\hbox{$\; \buildrel < \over \sim \;$}}
\def\simgt{\lower.5ex\hbox{$\; \buildrel > \over \sim \;$}}
\def\lum{erg s$^{-1}$}
\def\mbh{{$M_{\rm BH}$}}

\def\mnras{{MNRAS}}

\def\pasp{{PASP}}

\def\percm2{cm$^{-2}$}

\def\solum{$L_\odot$}
\def\solmass{$M_\odot$}

\def\oiii{[\ion{O}{3}]}

\def\feii{\ion{Fe}{2}}

\def\lbol{$L_{{\rm bol}}$}
\def\ledd{$L_{{\rm Edd}}$}
\def\lledd{$L_{{\rm bol}}/L_{{\rm Edd}}$}
\def\ser{S\'ersic}

\def\mbh{{$M_{\rm BH}$}}
\def\msig{{$M_{\rm BH}-\sigma_*$}}

\def\mmb{{$M_{\rm BH}-M_{\rm bulge}$}}

\slugcomment{To appear in
{\it The Astrophysical Journal}.}
\shorttitle{Black Hole Mass Scale in AGNs}
\shortauthors{HO \& KIM}

\begin{document}

\title{The Black Hole Mass Scale of Classical and Pseudo Bulges in Active 
Galaxies}

\author{Luis C. Ho\altaffilmark{1,2,3} and Minjin Kim\altaffilmark{3,4,5}}

\altaffiltext{1}{Kavli Institute for Astronomy and Astrophysics,
Peking University, Beijing 100871, China}

\altaffiltext{2}{Department of Astronomy,
Peking University, Beijing 100871, China}

\altaffiltext{3}{The Observatories of the Carnegie Institution for Science,
813 Santa Barbara Street, Pasadena, CA 91101, USA}

\altaffiltext{4}{Korea Astronomy and Space Science Institute, Daejeon 305-348,
Republic of Korea}

\altaffiltext{5}{KASI-Carnegie Fellow}

\begin{abstract}
The mass estimator used to calculate black hole (BH) masses in broad-line 
active galactic nuclei (AGNs) relies on a virial coefficient (the ``$f$ 
factor'') that is determined by comparing reverberation-mapped (RM) AGNs with 
measured bulge stellar velocity dispersions against the \msig\ relation of 
inactive galaxies.  It has recently been recognized that only classical bulges 
and ellipticals obey a tight \msig\ relation; pseudobulges have a different 
zero point and much larger scatter.  Motivated by these developments, we 
reevaluate the $f$ factor for RM AGNs with available $\sigma_*$ measurements, 
updated H$\beta$ RM lags, and new bulge classifications based on detailed 
decomposition of high-resolution ground-based and space-based images.  
Separate calibrations are provided for the two bulge types, whose virial 
coefficients differ by a factor of $\sim 2$: $f=6.3\pm1.5$ for classical 
bulges and ellipticals and $f = 3.2\pm0.7$ for pseudobulges.  The structure 
and kinematics of the broad-line region, at least as crudely encoded in the 
$f$ factor, seems to related to the large-scale properties or formation 
history of the bulge.  Lastly, we investigate the bulge stellar masses of the 
RM AGNs, show evidence for recent star formation in the AGN hosts that 
correlates with Eddington ratio, and discuss the potential utility of the 
\mmb\ relation as a more promising alternative to the conventionally used 
\msig\ relation for future refinement of the virial mass estimator for AGNs.
\end{abstract}

\keywords{galaxies: active --- galaxies: nuclei --- galaxies: Seyfert ---
quasars: emission lines --- quasars: general}

\section{Background}

The discovery of supermassive black holes (BHs) and empirical relations
between BH mass and host galaxy properties ushered an exciting era in 
extragalactic astronomy over the last decade and a half.  Particularly 
influential was the realization that BH mass correlates strongly with 
the stellar luminosity or mass (Kormendy \& Richstone 1995; Magorrian et al. 
1998) and velocity dispersion (Ferrarese \& Merritt 2000; Gebhardt et al. 
2000a) of the bulge.  Supermassive BHs are not only commonplace, but they may 
play a pivotal role in regulating many aspects of galaxy formation.  

Notwithstanding these advances, direct dynamical detections of supermassive 
BHs do not extend beyond the very nearby Universe (\lax 100 Mpc), and even 
then they are confined only to relatively inactive galaxies.  The bulk of the
growth of supermassive BHs occurred in more active systems, mostly at earlier 
epochs.  If we are to reach a fuller understanding of the interplay between 
BH accretion and galaxy evolution, we must devise a method to measure BH 
masses in active galactic nuclei (AGNs).

Historically, estimating BH masses in AGNs has always been challenging. Methods
that rely on continuum fitting are notoriously degenerate and model-dependent 
(e.g., Sun \& Malkan 1989; Sincell \& Krolik 1998).  Apart from limitations 
imposed by resolution, standard techniques involving spatially resolved 
kinematics of stars generally cannot be applied to AGNs because the strong 
featureless, nonstellar continuum severely dilutes the spectral features 
of the stars (e.g., Davies et al. 2006; Onken et al. 2007).  Low-density, 
narrow-line gas can be easily detected but its kinematics can be influenced by 
nongravitational effects (e.g., Hicks \& Malkan 2008).  Thus, it has long been 
realized that the dense, high-velocity gas that comprises the broad-line 
region (BLR) may offer the most promising probe of the gravitational potential 
of the BH.  If the gravity of the BH controls the motions of the BLR clouds, 
then from the virial theorem 

\begin{equation}
M_{\rm BH} = f{{r(\Delta V)^2}\over{G}} \equiv f {\rm VP},
\end{equation}

\noindent
where $r$ is the radius of the BLR, $\Delta V$ is the velocity width of the 
broad emission lines, $G$ is the gravitational constant, and VP is commonly   
called the virial product.  The coefficient $f$ depends on the kinematics, 
geometry, and inclination of the clouds and is likely to vary from object to 
object.  

Now, $r$, with scales of light days to light weeks (Kaspi et al. 2000), cannot 
be resolved spatially except in a few rare cases magnified by gravitational 
lensing (e.g., Sluse et al. 2012).  It can only be inferred indirectly.  Dibai 
(1977, 1984) pioneered the earliest efforts to constrain $r$  
through photoionization modeling of the BLR.  Subsequent reverberation mapping 
(RM; Blandford \& McKee 1982) of nearby Seyfert 1 galaxies showed that the 
photoionization method systematically overestimated $r$, by factors 
of $\sim$10 (e.g., Peterson 1993; but see recent improvements in 
photoionization estimates of $r$ by Negrete et al. 2013).  Despite 
the success of RM in improving the accuracy of BLR size measurements, and even 
though a statistically interesting sample of such measurements had already 
been accumulated (Ho 1999; Wandel et al. 1999), the community initially did 
not widely embrace the promise of this new technique.  Worries lingered over 
the key assumption of whether the BLR clouds are truly gravitationally bound 
(Krolik 2001).

The crucial breakthrough came with the discovery of the \msig\ relation in 
nearby inactive galaxies (Ferrarese \& Merritt 2000; Gebhardt et al. 2000a) 
and the demonstration that RM AGNs---at least the handful at the time that 
had stellar velocity dispersions measured---seem to obey an \msig\ 
relation that roughly parallels that of non-AGNs (Gebhardt et al. 2000b; 
Ferrarese et al. 2001; Nelson et al. 2004).  If, as can be reasonably supposed,
active and inactive galaxies follow the same \msig\ relation, then scaling 
the RM AGNs to the inactive galaxies yields an empirical calibration of the 
average value of the factor $f$ in Equation~1.  This is the approach adopted 
by Onken et al. (2004), Woo et al. (2010), Graham et al. (2011), Grier et al. 
(2013b), among others.

Two recent developments compel us to reevaluate the zero point of the AGN 
mass scale.  Both are reviewed by Kormendy \& Ho (2013) and in Section~2.  
Section~3 summarizes the most up-to-date database on the host galaxy 
properties of RM AGNs.  We utilize the latest \msig\ relation to calibrate 
$f$ (Section~4).  Implications and outstanding uncertainties are discussed in 
Section~5.  Section~6 ends with a summary and future prospects.

\section{The BH-Bulge Relations of Inactive Galaxies}

Kormendy \& Ho (2013) present a comprehensive review of the 
current status of BH mass measurements determined through dynamical modeling 
of spatially resolved kinematics.  Their database contains several 
improvements that lead to important quantitative revisions of the \msig\ and 
\mmb\ relations.  The key features are the following.

\begin{itemize}

\item{The sample, the largest and most comprehensive to date, contains 88 
galaxies. Apart from new detections, Kormendy \& Ho update a number of 
galaxies with recently improved stellar-dynamical BH masses.}

\item{The sample has been carefully scrutinized to identify three types of 
systematic problems that were previously underappreciated: (1) mass measurements
based on ionized gas kinematics that are biased low due to neglect of gas 
pressure support; (2) mergers-in-progress that have undermassive BHs; and (3) 
a few peculiar compact objects, either tidally stripped systems or remnants of 
high-redshift galaxies, that have overly massive BHs.  These outliers should 
not be included in the final correlations.}

\item{The database contains accurate bulge photometry based on a uniform set of 
bulge-to-disk decomposition of $K$-band images.  The luminosities are converted
to stellar masses using newly derived $K$-band mass-to-light ratios.  The 
database also presents consistently measured stellar velocity dispersions for 
the bulges.}

\item{Importantly, Kormendy \& Ho apply a self-consistent set of criteria 
to separate pseudobulges from classical bulges (Kormendy \& Kennicutt 2004), a 
distinction that turns out to be critical to their redefinition of the 
BH-host scaling relations.  Classical bulges and ellipticals define tight 
\mmb\ and \msig\ relations, whereas pseudobulges do not.  Kormendy \& Ho propose
that only classical bulges and ellipticals, both formed through gas-rich 
mergers at early times, participate in BH-galaxy coevolution.  Pseudobulges 
experience slow, stochastic growth via secular processes and do not coevolve 
closely with their central BH.}

\item{Excluding pseudobulges,

\begin{equation}
{{M_{\rm BH}} \over {10^9~M_\odot}} = \biggl(0.309^{+0.037}_{-0.033}\biggr)\
\biggl({{\sigma_*} \over {200~{\rm km~s}^{-1}}}\biggr)^{4.38 \pm 0.29}
\end{equation}

\noindent
and

\begin{equation}
{{M_{\rm BH}} \over {10^9~M_\odot}} = \biggl(0.49^{+0.06}_{-0.05}\biggr)\
\biggl({{M_{\rm bulge}} \over {10^{11}\,M_{\odot}}}\biggr)^{1.16 \pm 0.08}.
\end{equation}

\noindent
The intrinsic scatter is 0.29 dex for both relations: the \msig\ relation is 
{\it not}\ intrinsically tighter than the \mmb\ relation.  Equation~2 is 
slightly steeper than the best-fit relation of G\"ultekin et al. (2009), but 
not as extreme as that of McConnell \& Ma (2013) based on 
the entire sample that does not distinguish by bulge type.   Equation~3 is 
marginally steeper than the near-linear relation of H\"aring \& Rix (2004), 
but the most important difference lies in the normalization, which is on 
average $\sim 4$ times higher than the canonical value of $\sim 0.13$
(Kormendy \& Gebhardt 2001; H\"aring \& Rix 2004).
}
\end{itemize}

\section{Bulge Properties and Virial Products for the RM Sample}

In light of the developments discussed in Section~2, the factor $f$ for the RM 
AGN sample should be reevaluated taking into account the bulge type of the 
host galaxies.  Our approach is fundamentally different from that of other 
recent efforts to quantify $f$ (Woo et al. 2010, 2013; Park et al. 2012a; Grier
et al. 2013b), which do not explicitly distinguish between pseudo and classical
bulges.  The work of Graham et al. (2011) comes closest to the spirit of this 
paper, but it primarily emphasizes the role of bars, which, as Kormendy \& Ho 
emphasize, is related to, but not synonymous with, pseudobulges.  

We devote considerable attention to classify the bulge type of the RM AGN 
sample in a consistent manner.  The criteria used to identify pseudobulges 
(Kormendy \& Kennicutt 2004) are summarized and augmented in Kormendy \& Ho 
(2013; Supplemental Material).  However, in practice some of the criteria 
normally used to classify inactive galaxies are difficult to apply to type~1 
(broad-line) Seyferts and quasars, whose host galaxies, especially their 
bulges, are often severely contaminated by the bright active nucleus.  
Appendix~A gives notes for each object in our sample, which was individually 
scrutinized using all available photometric information in the literature.  We 
give preference to two-dimensional structural decomposition of high-resolution 
images---principally, but not exclusively, based on {\it Hubble Space 
Telescope (HST)}\ observations---with sufficient detail to extract a robust 
measurement of the bulge-to-total ($B/T$) luminosity ratio and a bulge surface 
brightness distribution parameterized by the \ser\ (1968) function.  Recently,
Grier et al. (2013a) also attempted to assign bulge types to the RM sample,
but we disagree with a number of their classifications.

Following common practice (Kormendy \& Kennicutt 2004; Fisher \& Drory 2008),
we consider bulges with a \ser\ index $n < 2$ as pseudobulges and those with 
$n \ge 2$ as classical bulges.  However, as illustrated in Appendix~A, a 
practical challenge arises from the fact that it is often difficult to 
decompose the bulge accurately in AGN hosts, let alone a higher order 
attribute such as its profile shape.   The luminosity of the bulge is a more 
robust quantity.  Thus, apart from the \ser\ index, we also take into 
consideration the relative light fraction of the bulge, generally favoring a 
pseudobulge classification for systems with 

\begin{figure*}[t]
\centerline{\psfig{file=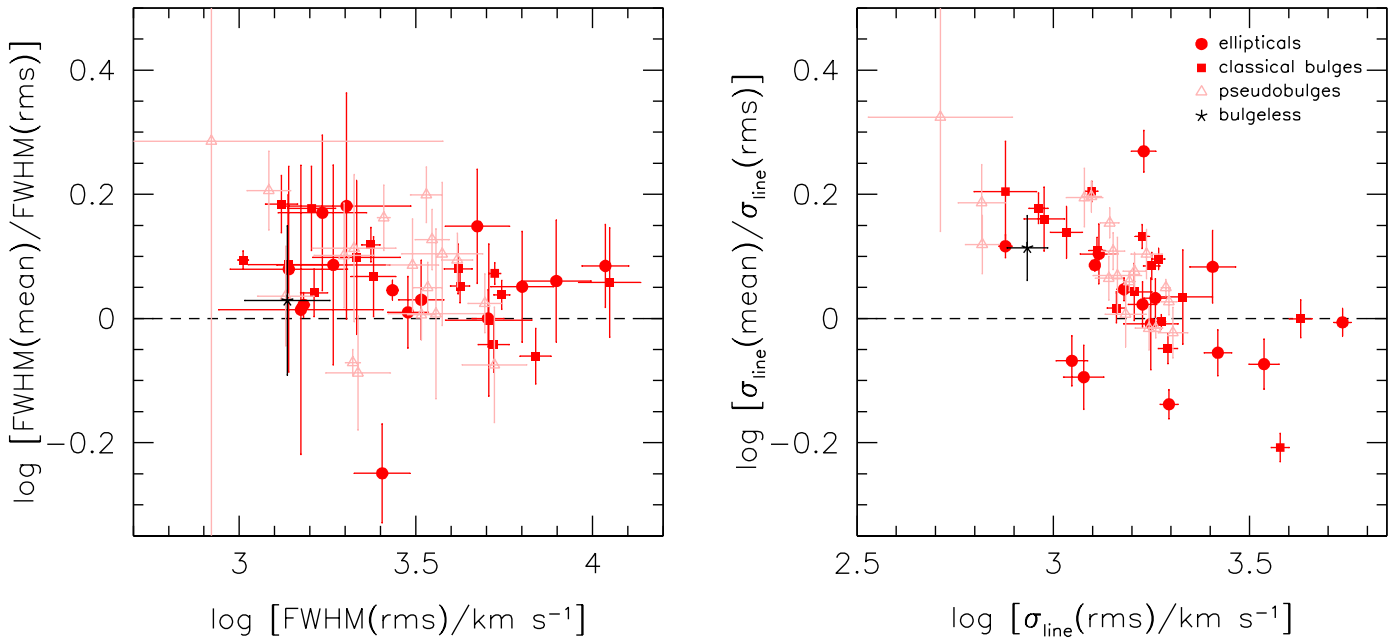,width=17.5cm,angle=270}}
\figcaption[fig1.ps]{Comparison of (left) FWHM and (right) $\sigma_{\rm line}$
of \hb\ measured from mean and rms spectra.  The line widths from the mean
spectra are systematically larger, by $\sim 15\%-16\%$.  
\label{fig1}}
\end{figure*}

\noindent
$B/T$ \lax\ 0.2 (Fisher \& Drory 
2008; Gadotti 2009).  In some cases, additional clues come from the detection 
of circumnuclear rings and other signatures of ongoing central star formation.  

Our RM AGN sample consists of the 44 objects that has been imaged, to varying 
degrees of accuracy, at rest-frame optical wavelengths with \hst.  This 
comprises the majority ($\sim 90\%$) of all sources with published RM data.
Appendix~A describes details relevant to our bulge classification, and Table~1 
summarizes the final results.  Among the 44 objects, we formally classify 13 
as ellipticals, 14 as classical bulges, and 16 as pseudobulges; one 
(PG~1022+519 or Mrk~142) does not have a detectable bulge component (Bentz et 
al. 2013) and hence cannot be classified.  Reliable measurements of central 
stellar velocity dispersion are available for a total of 31 sources, among 
them 15 classical bulges or ellipticals and all of the 16 pseudobulges.  For 
all 43 of the galaxies with detected bulges, we convert their bulge 
luminosities to the $R$ band following the procedure of Kim et al. (2008a) and 
Kim et al. (2014; hereinafter K14).  This involves transforming each of the 
observed \hst\ filters to $R$ assuming the fiducial composite quasar spectrum
from Vanden~Berk et al. (2001) for the nucleus and the galaxy spectral 
templates from Calzetti et al. (1994) and Kinney et al. (1996) for the bulge.  
The choice of galaxy template is based on the Hubble type inferred from the 
measured $B/T$, using the statistical correlation between the morphological 
type and $B/T$ of Simien \& de~Vaucouleurs (1986).

We calculate virial products using improved time lag measurements based on the 
SPEAR method of Zu et al. (2011), as advocated by Grier et al. (2013b).  
Whenever possible we adopt the revised time lags from the recent compilation 
of Grier et al. (2013b).  The majority of those not included in that work have 
revised lags given in Zu et al. Of the 43 sources with bulge parameters, 39 
(91\%) have revised and improved H$\beta$ lags (Table~2).

With the notable exception of Collin et al. (2006), previous works calculate 
$f$ using the virial product based on the dispersion (second moment) of 
H$\beta$, $\sigma_{\rm line}$, measured from the root-mean-square (rms) 
spectrum.  While there are well-motivated reasons for this choice (e.g., 
Peterson et al. 2004), a priori it is not obvious that $\sigma_{\rm line}$ 
leads to a better calibration of $f$ than FWHM, although Collin et al. (2006) 
argue that $\sigma_{\rm line}$ is a less biased estimator of $\Delta V$ than 
FWHM. Moreover, rms and mean spectra may yield systematically different line 
widths.  Table~2 lists line widths measured from both rms and mean spectra and 
their respective virial products.  Published line widths measured from mean 
spectra exist for all but five of the 44 RM AGNs, including 26 of the 31 
sources with $\sigma_*$.  Consistent with past studies (e.g., Sergeev 
et al. 1999; Collin et al. 2006), we confirm that mean spectra yield 
systematically broader H$\beta$ line widths than rms spectra (Figure~1), on 
average by $\sim 16\%$ ($0.063\pm0.089$ dex for FWHM and $0.064\pm0.101$ for 
$\sigma_{\rm line}$).  Curiously, the discrepancy between mean and rms 
velocities is most severe for systems with $\sigma_{\rm line}$ \lax\ 2000 
\kms, and the amount of departure increases systematically with decreasing 
$\sigma_{\rm line}$ (right panel of Figure~1).  Park et al. (2012b) noticed 
this effect in a significantly smaller sample.  The trend also appears for 
FWHM but it is less well defined; FWHM(mean) statistically exceeds FWHM(rms) 
at all velocities, although the largest excess occurs at FWHM \lax\ 3000 \kms\ 
(left panel of Figure~1).

\section{Calibrating $\lowercase{f}$}

As in previous works, we compute $\langle f \rangle$ by assuming that the RM 
AGNs intrinsically obey the same \msig\ relation as inactive galaxies.  The 
key difference between our approach and that of past studies is that we 
initially restrict the calibration 

\clearpage
\begin{figure*}[t]
\centerline{\psfig{file=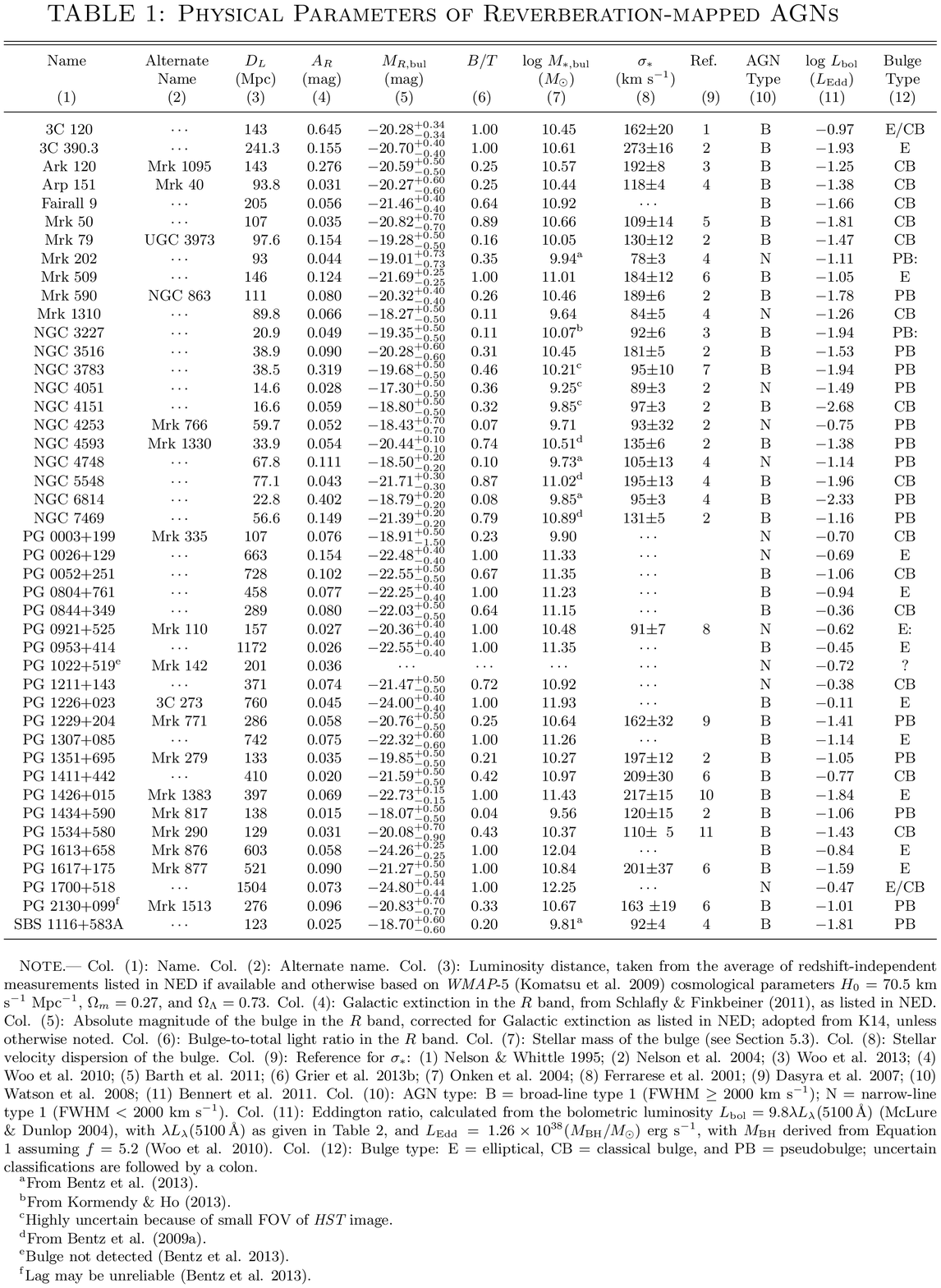,width=7.3in,angle=0}}
\end{figure*}

\clearpage
\begin{figure*}[t]
\centerline{\psfig{file=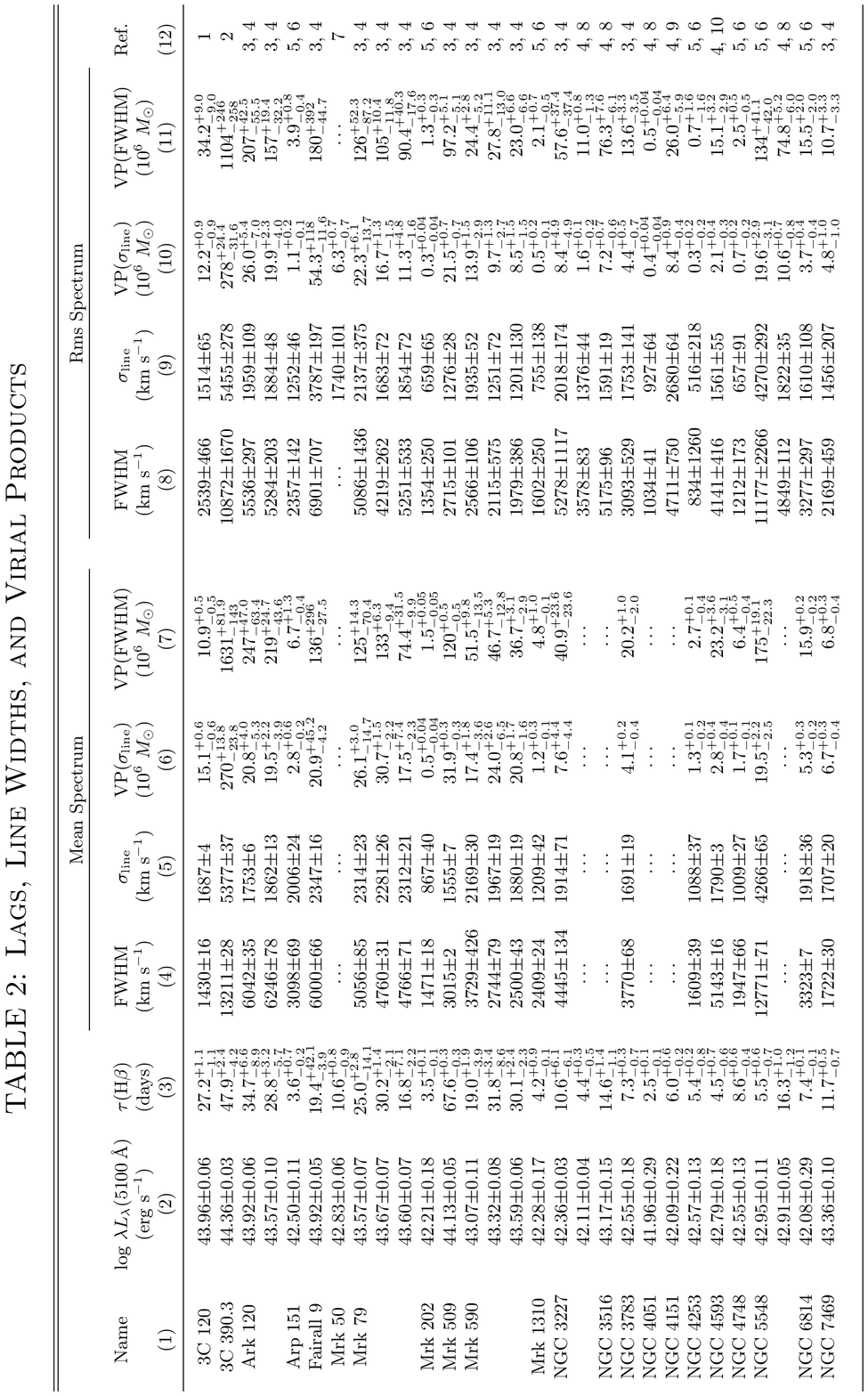,height=10in,angle=180}}
\end{figure*}

\clearpage
\begin{figure*}[t]
\centerline{\psfig{file=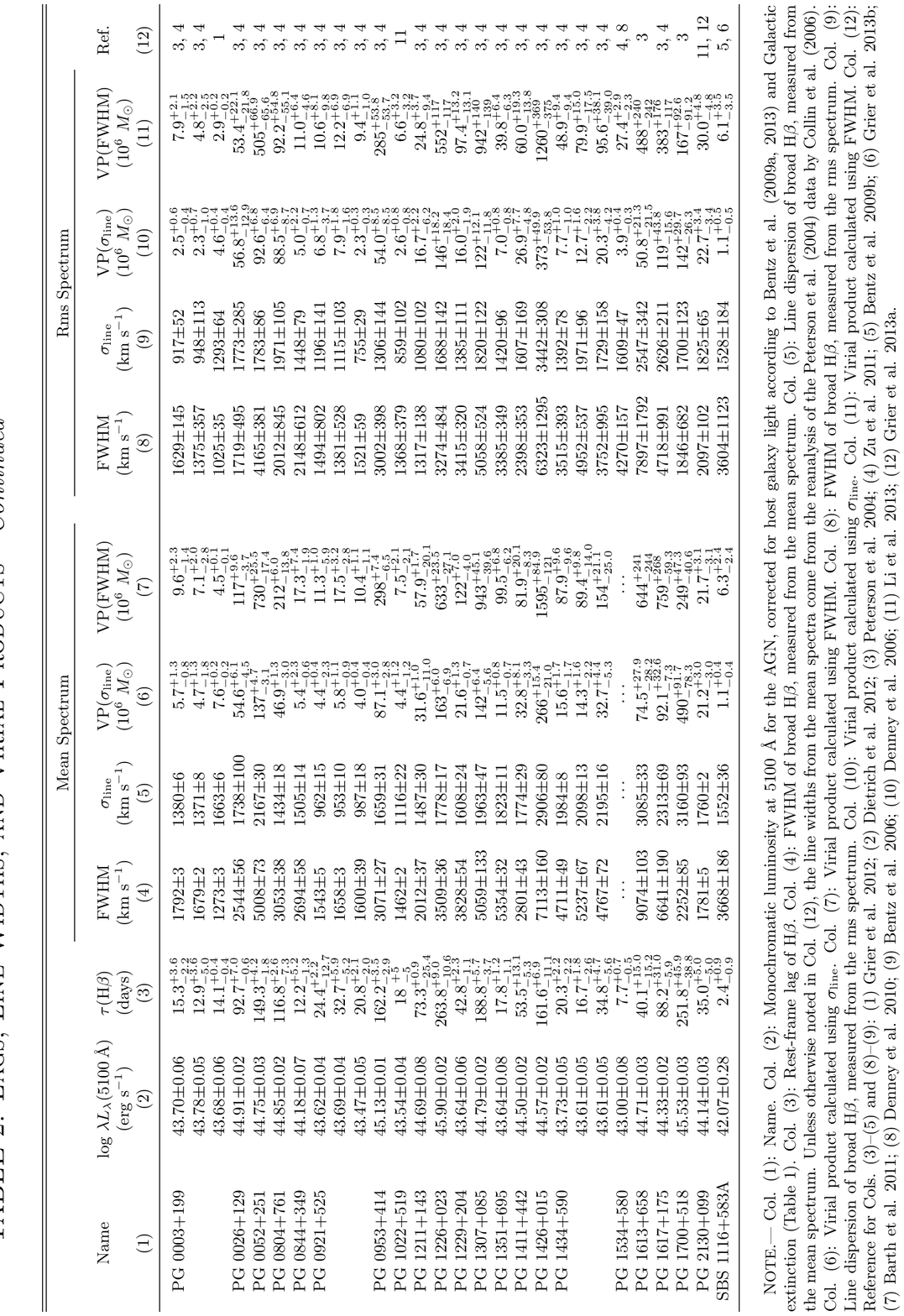,height=10in,angle=180}}
\end{figure*}
\clearpage
\vskip 0.3cm
\psfig{file=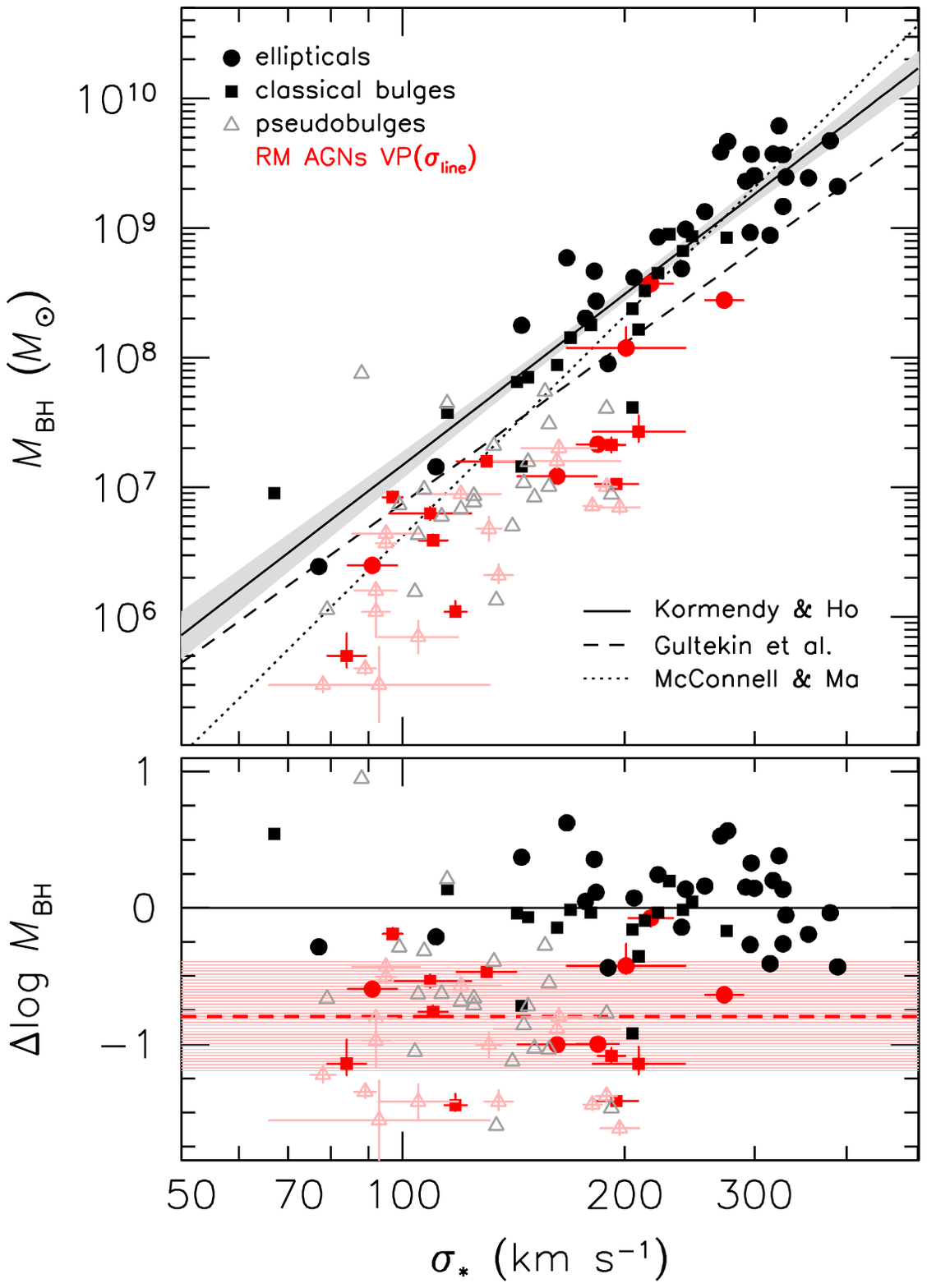,width=8.75cm,angle=0}
\figcaption[fig2.ps]{
({\it Top}) The \msig\ relation for inactive galaxies (black points) and RM
AGNs (red points).  The masses for the RM AGNs represent the virial product,
${\rm VP} = c \tau \Delta V^2/G \equiv M_{\rm BH}/f$, with $\Delta V =
\sigma_{\rm line}$(\hb) measured from rms spectra.  Classical bulges and
ellipticals are highlighted as filled symbols, and pseudobulges are plotted as
open symbols.  Error bars are suppressed for the inactive galaxies to reduce
crowding.  The best-fit relation of Kormendy \& Ho for classical bulges and
ellipticals (Equation~2) is given by the solid line; the gray shading represents
its $1\sigma$ scatter.  The fits of G\"ultekin et al. (2009) and McConnell \&
Ma (2013) are shown as dashed and dotted lines, respectively.  ({\it Bottom})
Residuals of the data points with respect to the Kormendy \& Ho fit for
ellipticals and classical bulges. The red dashed line and associated shaded
pink band mark the average offset and standard deviation for the 15 RM AGNs
hosted by ellipticals and classical bulges: $\langle \Delta \log M_{\rm BH}
\rangle=-0.79\pm0.42$.
\label{fig2}}
\vskip 0.3cm

\noindent
exclusively to classical bulges and 
ellipticals, which Kormendy \& Ho (2013) show to be the only subgroup of local 
inactive galaxies that obey a tight \msig\ relation.  Pseudobulges do not.

Figure~2 shows the \msig\ distribution for inactive galaxies (solid black
points) with spatially resolved dynamical BH mass measurements from the 
database of Kormendy \& Ho, with their best-fitting relation (Equation~2) for 
classical bulges and ellipticals overplotted as a solid line accompanied by its 
1$\sigma$ range in gray shading.  Pseudobulges, excluded from the fit, are 
shown as open symbols in a lighter shade. For comparison, we also show the 
G\"ultekin et al. (2009) and McConnell \& Ma (2013) fits, which include both
bulge types mixed.  The colored points represent the subset of 31 RM AGNs that 
have measurements of $\sigma_*$; plotted on the ordinate is the virial product 
${\rm VP} \equiv M_{\rm BH}/f$ based on $\Delta V$ parameterized through 
$\sigma_{\rm line}$(\hb) measured from rms spectra.  The bottom panel shows 
the residuals of the data points with respect to the best-fit relation.  On 
average, the RM AGNs in classical bulges and ellipticals are offset from their
inactive counterparts by 0.79 dex (factor 6.2; red dashed line), 

\vskip 0.3cm
\psfig{file=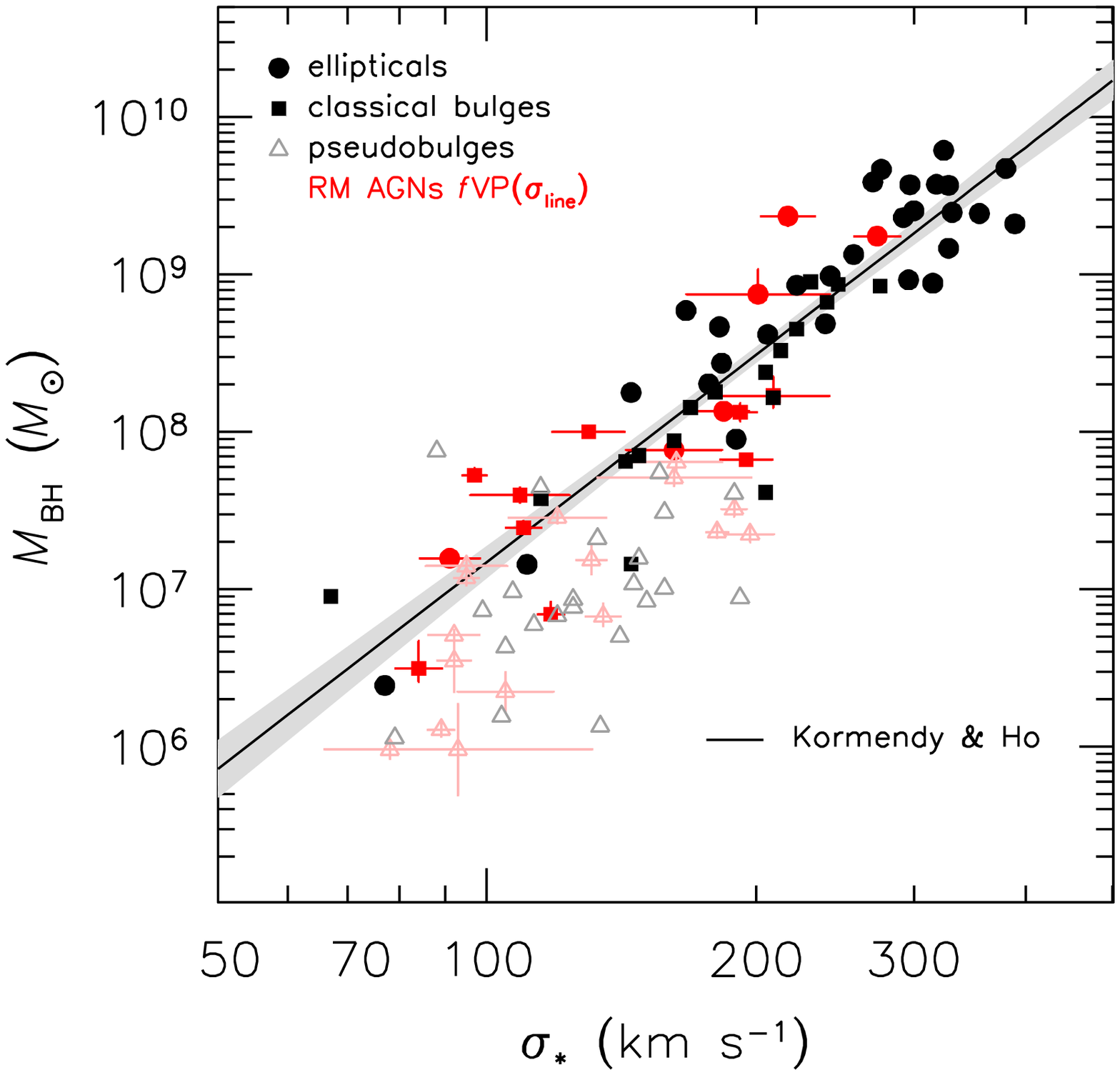,width=8.75cm,angle=0}
\figcaption[fig3.ps]{The \msig\ relation for RM AGNs (red points) with the 
virial products scaled by $f$ to match the locus of inactive galaxies (black 
points).  Error bars are suppressed for the inactive galaxies to reduce 
crowding.  The virial products, derived from $\sigma_{\rm line}$(\hb) measured 
from rms spectra, have been scaled by $f = 6.3$ for classical bulges and 
ellipticals and by $f = 3.2$ for pseudobulges.  The best-fit relation of 
Kormendy \& Ho for classical bulges and ellipticals (Equation~2) is given by 
the solid line; the gray shading denotes its $1\sigma$ scatter.
\label{fig3}}
\vskip 0.3cm

\noindent
with a
standard deviation of 0.42 dex (pink shaded band).  This is a zeroth-order
estimate of $f$.

To obtain a more formal determination of $f$ and its intrinsic scatter, we 
express the \msig\ relation as

\begin{equation}
\log \biggl( {{f {\rm VP}} \over {M_\odot}}\biggr) = \alpha + \beta \, \log 
\biggl({{\sigma_*}\over{200\,{\rm km\,s}^{-1}}}\biggr),
\end{equation}

\noindent
where $\alpha$ is the normalization and $\beta$ is the slope.  We use the 
{\tt FITEXY} estimator (Press et al. 1992), as modified by Tremaine et al. 
(2002), to perform a linear regression to minimize the quantity 

\begin{equation}
\chi^2 = \sum_{i=1}^{N} \frac{(y_i + \log f - \alpha - \beta x_i)^2}
{\epsilon^2_{yi} + \beta^2 \epsilon^2_{xi} + \epsilon^2_{0}}.
\end{equation}

\noindent
Park et al. (2012a) discuss at length the performance of alternate fitting 
methods.  In the above expression, $\alpha$ and $\beta$ are the regression 
coefficients, $y = \log ({\rm VP}/M_\odot)$, $x = \log (\sigma_*/200\,
{\rm km\,s}^{-1})$, $\epsilon_y$ and $\epsilon_x$ are their respective 
measurement errors, and $\epsilon_0$ is the intrinsic scatter.  {\tt FITEXY} 
treats $x$ and $y$ as independent variables and allows for asymmetric errors.  
Fixing the intercept ($\alpha=8.49$) and slope ($\beta = 4.38$) to those of 
the \msig\ relation for inactive galaxies, we solve for the two parameters $f$ 
and $\epsilon_0$.  For VP based on $\sigma_{\rm line}$(H$\beta$), we find 
$f=6.3\pm1.5$ and $\epsilon_0 = 0.39\pm0.07$ dex, consistent with the simple 
estimate above.  In the case of VP based on FWHM(H$\beta$), $f = 1.5\pm 0.4$.  
For completeness, Table~3 also gives fits for VP calculated from mean spectra 
instead of rms spectra.  The errors on the parameters come from a bootstrap 
calculation.  We perform the same calculation using randomly selected 

\begin{figure*}[t]
\centerline{\psfig{file=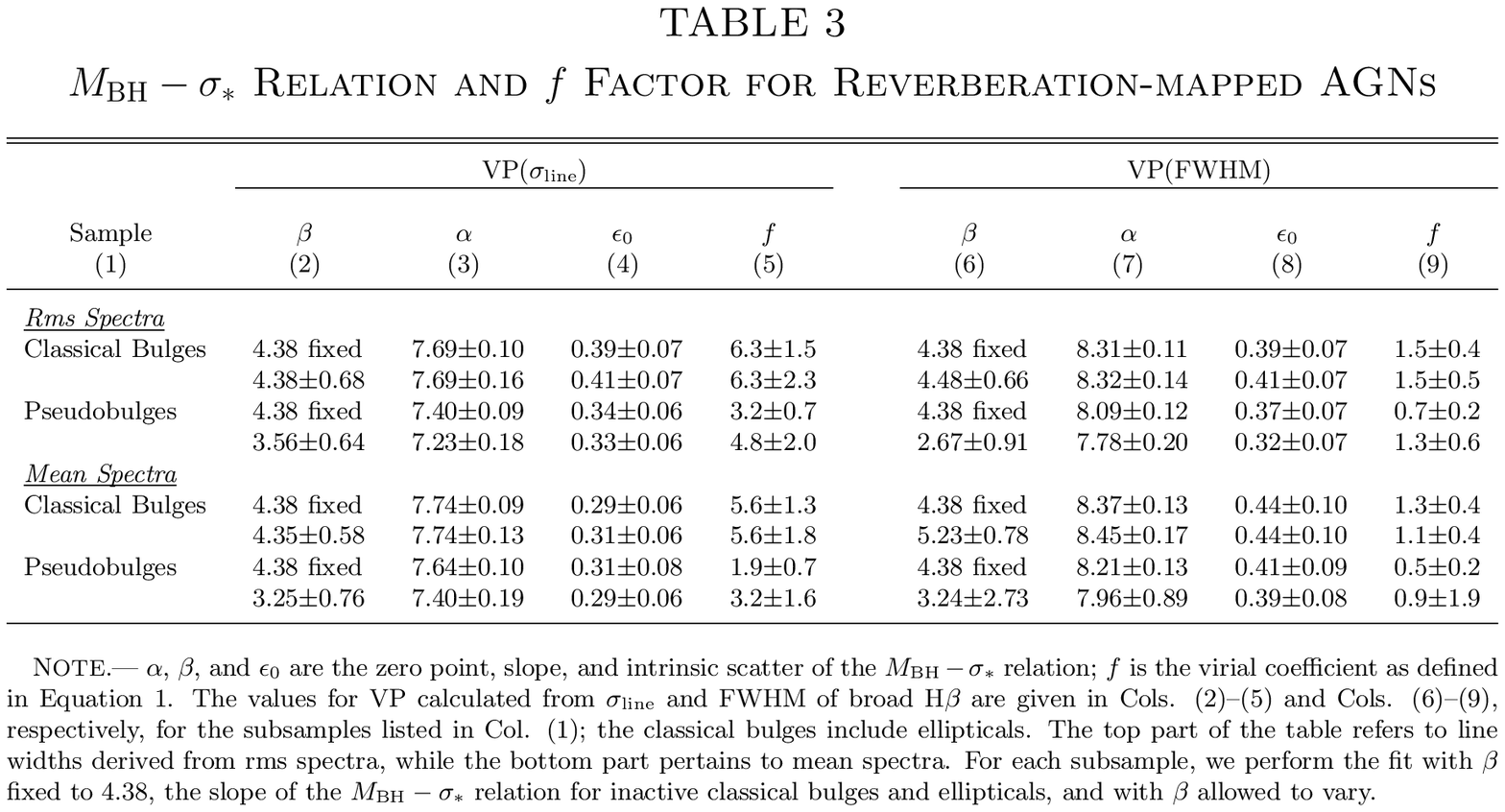,width=7.5in,angle=0}}
\end{figure*}

\noindent
subsamples repeated 5000 times. The errors are determined by the 
standard deviation of the best parameters derived from the repeated estimation.

The above estimate of $f$ strictly holds only for AGNs hosted by elliptical 
galaxies and classical bulges.  We cannot apply the same calibration strategy 
to systems containing pseudobulges because their inactive counterparts exhibit 
a significantly different (Hu 2008) and much looser (Greene et al. 
2010; Kormendy et al.  2011; Kormendy \& Ho 2013) \msig\ relation.  Although 
the scatter of the \msig\ relation for pseudobulges is indeed large, Figure~2 
illustrates that, as a group, pseudobulges tend to lie systematically below 
the locus of classical bulges and ellipticals.  Kormendy \& Ho resisted from 
performing a formal fit to the pseudobulges on account of the questionable 
physical significance of the \msig\ relation for these objects.   Here, we have
a more pragmatic concern: without a reference point for pseudobulges, even if 
imprecise, we lose all ability to estimate BH masses for a large segment of 
the AGN population, many of which have modest \mbh\ that likely live in disk 
host galaxies with pseudobulges.  We proceed as follows.

Keeping the slope fixed to $\beta = 4.38$, the pseudobulge population 
(tabulated in Table~3 of Kormendy \& Ho) has a zero point of $\alpha = 
7.91\pm0.11$ and an intrinsic scatter of $\epsilon_0 = 0.46\pm0.15$ dex.  If 
we {\it assume}\ that inactive pseudobulges follow an \msig\ relation with the 
same slope as classical bulges but with a zero point offset by this 
magnitude\footnote{A formal fit to the pseudobulges alone yields a 
significantly shallower but highly uncertain slope of $\beta = 2.66\pm1.61$, 
$\alpha=7.54\pm0.30$, and $\epsilon_0 = 0.44\pm0.12$ dex.}, then we can derive 
a rough estimate of the corresponding $f$ factor for the 16 RM AGNs hosted by 
pseudobulges.  Following the same procedure as above, the {\tt FITEXY} 
estimator yields $f = 3.2\pm0.7$ for VP($\sigma_{\rm line}$) and $f = 
0.7\pm0.2$ for VP(FWHM).  AGNs in classical bulges and pseudobulges do 
{\it not}\ share the same value of $f$; the former has a value $\sim 2$ 
times larger.  Figure~3 redisplays the \msig\ relation for RM AGNs, now with 
the virial products properly scaled by the bulge type-specific $f$ factor.  
The AGN hosts---by design---faithfully track their inactive counterparts.

The above results were obtained by fixing $\beta$ to 4.38, the slope of the 
\msig\ relation for inactive classical bulges.  It is interesting to note that 
an unconstrained fit on VP($\sigma_{\rm line}$) for the RM AGNs yields 
$\beta = 4.38\pm0.68$ and $3.56\pm0.64$ for the classical and pseudo bulge 
subsamples, respectively.  Both slopes are consistent with those of their 
respective counterparts in the inactive galaxy sample.  The \msig\ relation 
of pseudobulges---whether their BHs are active or not---is marginally 
shallower than that of classical bulges.  For completeness, Table 3 lists 
$f$ factors inferred for fits with $\beta$ unconstrained, but for these cases 
the uncertainties are large and we recommend against them for actual 
applications. 

\section{Discussion}

\subsection{The $f$ Factor Depends on Bulge Type}

A number of previous works have used the \msig\ relation of inactive galaxies 
to derive the zero point ($f$ factor) of the virial mass estimator for AGNs 
(Equation~1).  Onken et al. (2004) first applied this calibration, referenced 
with respect to the \msig\ relation of quiescent galaxies of Tremaine et al. 
(2002), to a subset of 14 RM AGNs with available measurements of $\sigma_*$ 
to arrive at $f = 5.5\pm1.7$.  This value of $f$ pertains to VP calculated 
from $\sigma_{\rm line}$(\hb).  Woo et al. (2010) enlarged the sample to 24 
objects and, using the G\"ultekin et al. (2009) \msig\ relation as reference, 
redetermined $f = 5.2\pm1.2$ and an intrinsic scatter of $\epsilon_0 = 
0.44\pm0.07$ dex.  Slight variants of this exercise, with essentially 
comparable results, appear in Park et al. (2012a) and Woo et al. (2013).  
Grier et al. (2013b) present the largest and latest collection of RM lags and 
stellar velocity dispersion measurements; using the quiescent galaxy sample of 
McConnell et al. (2011) as reference, they find $f = 4.19\pm1.08$ and no 
noticeable dependence on mass or galaxy morphology.  The most notable 
exception comes from Graham et al. (2011), who report a significantly lower 
value of $f = 2.8^{+0.7}_{-0.5}$; Park et al. (2012a) conclude that this large 
difference comes from sample selection and choice of regression method.

We take a different approach.  Motivated by the recent evidence that the 
BH-host scaling relations depend strongly on bulge type (Kormendy \& Ho 2013), 
we calibrate the $f$ factor 

\vskip 0.3cm
\begin{figure*}[t]
\centerline{\psfig{file=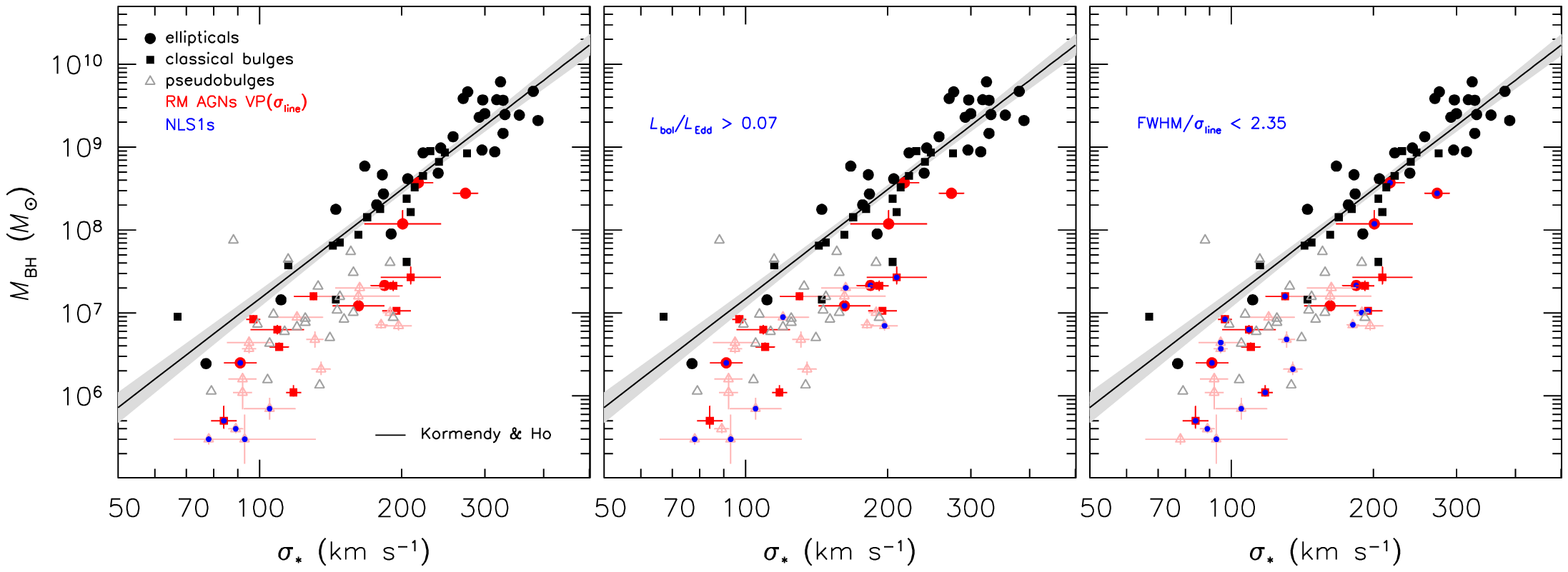,width=17.5cm,angle=270}}
\figcaption[fig4.ps]{
\msig\ relation for RM AGNs, as in Figure 2, highlighting (in blue) sources
classified as (left) NLS1s, (middle) those with Eddington ratios \lbol/\ledd\
$>$ 0.07, and (right) those with \hb\ core profiles narrower than a Gaussian,
FWHM/$\sigma_{\rm line} < 2.35$.
\label{fig4}}
\end{figure*}
\vskip 0.3cm

\noindent
separately for classical and pseudo bulges.  This
is by no means straightforward, in view of the vagaries of bulge 
classification, which is made all the more challenging in the presence of a 
bright active nucleus.  Nevertheless, we demonstrate that active galaxies, as 
their inactive counterparts, separately do define their own \msig\ relation 
according to bulge type.   Moreover, the slopes of the respective relations 
for classical and pseudo bulges are---within the considerable error 
bars---consistent between active and inactive galaxies.  Our unconstrained 
fits for VP($\sigma_{\rm line}$) based on rms spectra yield $\beta = 
4.38\pm0.68$ for classical bulges and $\beta = 3.56\pm0.64$ for pseudobulges.
Qualitatively similar differences between classical and pseudo bulges are 
found for VP(FWHM) and for line widths measured from mean spectra (see 
Table~3).  A number of previous studies (e.g., Greene \& Ho 2006; Woo et al. 
2010, 2013; Bennert et al. 2011; Xiao et al. 2011) conclude that the \msig\ 
relation of active galaxies tends to be shallower than that of inactive 
systems.  In light of the present findings, the previous results can be 
understood to be a natural consequence of the fact that current AGN samples 
with $\sigma_*$ measurements tend to contain a large proportion of later type 
galaxies, which preferentially host pseudobulges (Kormendy \& Kennicutt 2004).

Of more immediate relevance to the BH mass scale of AGNs, we show that AGNs 
hosted in pseudobulges lie slightly but systematically below the \msig\ 
relation compared to those hosted in classical bulges.  Again, this parallels 
the trend seen in inactive galaxies (Hu 2008; Greene et al. 2010; Kormendy et 
al. 2011; Kormendy \& Ho 2013), as Figure~2 clearly illustrates.  This being 
the case, the $f$ factor should depend on bulge type.  Indeed, with the slope 
of the \msig\ relation fixed to $\alpha = 4.38$, classical bulges and 
ellipticals on average have $f=6.3\pm1.5$, twice the value of $f =
3.2\pm0.7$ for pseudobulges.  These values pertain to VP($\sigma_{\rm line}$) 
derived from rms spectra, but the same trend holds for other measures of VP 
(Table~3).  Given that uncertainties in the $f$ factor currently dominate the 
overall error budget of BH virial masses, this factor-of-two difference in 
$f$ between classical and pseudo bulges represents a major source of 
systematic uncertainty that can---and should---be eliminated.  In practice, 
however, this will be challenging to implement.  Classification of the bulge 
type depends on detailed information on the structural parameters of the host 
galaxy, which is difficult to ascertain for distant or very luminous quasars 
wherein the bright nucleus overwhelms the signal from the central regions of 
the host.  Fortunately, pseudobulges tend to be confined to lower mass 
systems: most are characterized by $M_{\rm BH}$ \lax\ $10^8$ \solmass\ and 
$\sigma_*$ \lax\ 200 \kms\ (Figure~2).  For the foreseeable future, 
studies of high-redshift AGNs will be sensitive mostly to the upper end of the 
BH mass function, a regime likely to be dominated by ellipticals or 
early-type galaxies with classical bulges.  On the other hand, lower redshift 
AGNs with $M_{\rm BH} \approx 10^6-10^8$ \solmass\ will suffer from this 
factor-of-two uncertainty in $f$ unless the bulge type can be determined 
independently; pseudo and classical bulges thoroughly overlap within this 
mass range (Figure~2).   Only for $M_{\rm BH}$ \lax\ $10^6$ \solmass\ are AGN 
samples likely to be dominated by pseudobulges (e.g., Greene et al. 2008; 
Jiang et al. 2011).  In such low-mass systems, it would be relatively safe 
to adopt a priori the $f$ factor appropriate for pseudobulges.

Graham et al. (2011), studying the dependence of the \msig\ relation on bars,
show for the RM AGN sample that the $f$ factor for barred hosts is $2-3$ times 
lower than that for unbarred hosts.  Although the presence of a bar does not 
correlate perfectly with bulge type, the systematic difference in $f$ between 
barred and unbarred galaxies qualitatively resembles the dependence on bulge 
type that we find.

Taken at face value, the different $f$ factors for pseudo and classical bulges
suggest that the two groups on average have systematically different BLR
structures.  A value of $f \approx 3$ for pseudobulges formally corresponds to
a spherical BLR (Kaspi et al. 2000), whereas $f = 6.3$ for classical bulges
agrees better with a disk-like geometry.  It is unclear why the structure of
the BLR would depend on the large-scale properties of the host galaxy bulge.
From dynamical modeling of velocity-resolved RM data, Brewer et al. (2011)
infer $f=2.5\pm1.6$ for Arp~151, and Pancoast et al. (2012) obtain $f=6.0
\pm4.9$ for Mrk~50.  Both objects host a classical bulge (Table~1), although
the classification for Arp~151 is quite uncertain because of its strongly 
tidally distorted morphology (Appendix~A). We should also keep in 

\vskip 0.3cm
\psfig{file=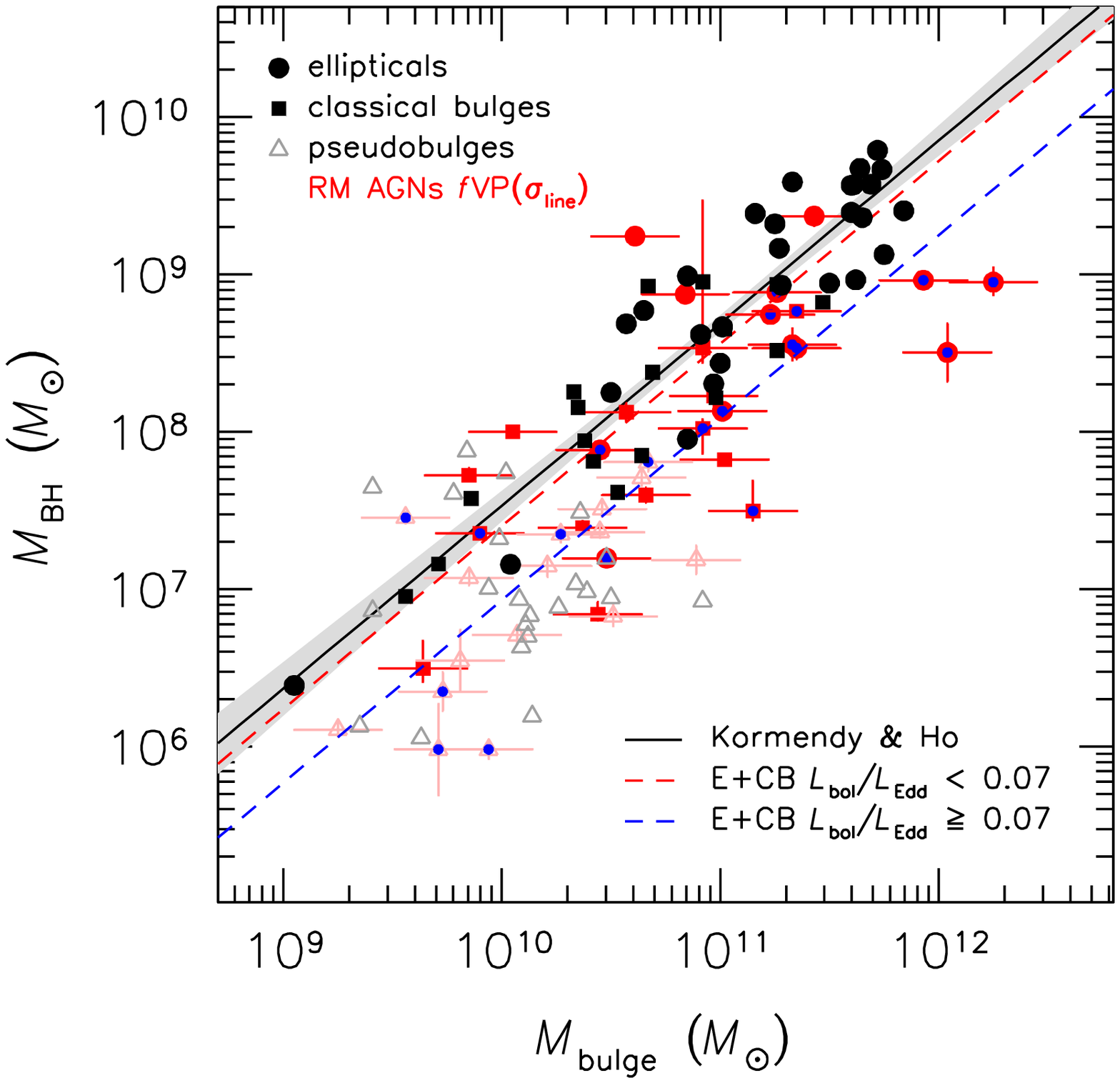,width=8.75cm,angle=0}
\figcaption[fig5.ps]{\mmb\ relation for inactive galaxies (black points) and
RM AGNs (red points).  Error bars are suppressed for the inactive galaxies to
reduce crowding.  The virial products, derived from $\sigma_{\rm line}$(\hb)
measured from rms spectra, have been scaled by $f = 6.3$ for classical bulges
and ellipticals and by $f = 3.2$ for pseudobulges.  The best-fit relation of
Kormendy \& Ho for classical bulges and ellipticals (Equation~3) is given by
the solid line; the gray shading denotes its $1\sigma$ scatter.  AGNs hosted
by classical bulges and ellipticals with \lbol/\ledd\ $<$ 0.07 are denoted by
filled red symbols and the red dashed line; those with \lbol/\ledd\ $\ge$ 0.07
are highlighted with a blue center and the blue dotted line.
\label{fig5}}
\vskip 0.3cm

\noindent
mind that 
the apparent difference in $f$ between pseudo and classical bulges can arise, 
at least in part, from uncertainties in bulge classification.

\subsection{Possible Dependence of $f$ on Eddington Ratio}

Figure~4 shows that the $f$ factor may also depend mildly on Eddington ratio.
The statistics are poor, but the evidence is suggestive.  Objects 
crudely\footnote{We do not have access to other additional commonly used 
criteria that include \feii\ and \oiii\ \lamb5007 strength.} classified as 
narrow-line Seyfert 1 galaxies (NLS1s; blue points in left panel) according 
to FWHM(H$\beta$) $<$ 2000 \kms, typically regarded as highly accreting 
systems (e.g., Boroson 2002), lie systematically offset toward lower \mbh\ at 
a given $\sigma_*$ than those with larger line widths.  The middle panel 
separates the objects by Eddington ratio to illustrate this effect more 
clearly.  The RM sample has a median Eddington ratio of \lledd\ = 0.07, 
calculated from the bolometric luminosity $L_{\rm bol} = 
9.8 \lambda L_\lambda$(5100 \AA) (McLure \& Dunlop 2004), with 
$\lambda  L_\lambda$(5100 \AA) given in Table~2, $L_{\rm Edd}\,=\,
1.26 \times 10^{38} (M_{\rm BH}/M_{\odot})$ \lum, and $M_{\rm BH}$ derived 
assuming a fiducial $f = 5.2$ from Woo et al. (2010).  Splitting the sample in 
half at this median value, objects with larger \lledd\ (blue points) lie 
displaced toward lower \mbh\ at a given $\sigma_*$ than those with smaller 
\lledd.  The magnitude of the offset is $\sim 0.2$ dex in log~\mbh, or, 
equivalently, in $\log f$.  The trend seems to hold separately for both 
pseudobulges and classical bulges alike.  

Collin et al. (2006) note that the shape of the \hb\ profile, as gauged by 
FWHM/$\sigma_{\rm line}$, varies systematically with line width and Eddington 
ratio.  Sources with NLS1-like velocity widths and high \lledd\ tend to have 
\hb\ profiles with narrower cores, which these authors suggest arise, at least 
in part, from inclination effects. They further argue that this systematic 
effect can be mitigated by using $\sigma_{\rm line}$ instead of 
FWHM when computing VP.  Thus, it appears unlikely that the effect seen in 
Figure~4 can be attributed to inclination alone.  Indeed, the right panel of 
the figure shows that the distribution of points does not depend on 
FWHM/$\sigma_{\rm line}$.  Interpreted literally, the structure and kinematics 
of the BLR, as crudely imprinted in the $f$ factor, correlate with Eddington 
ratio: higher accretion rates result in larger values of $f$.

\subsection{Evidence of Young Stars in AGN Hosts}

As a by-product of the structural decomposition of the host galaxies needed 
for classifying the bulge types, we have a uniform set of accurately determined
$R$-band luminosities for the bulge.  Unfortunately, we do not yet have 
sufficient color information to transform the luminosities into reliable 
stellar masses.  As an initial estimate, we assume $\log (M/L_R) =
-0.523 + 0.683(B-R)$ (Bell et al. 2003), which, for $B-R$ = 1.57 mag 
appropriate for an evolved elliptical galaxy (Fukugita et al. 1995), implies 
$M/L_R = 3.55\, M_\odot/L_\odot$.  We adopt $M_{\odot, R} = 4.46$ mag for the 
Sun\footnote{{\tt mips.as.arizona.edu/$\sim$cnaw/sun.html}}.

Figure~5 shows the \mmb\ relation for inactive galaxies from Kormendy \& Ho 
and for the 43 RM AGNs with bulge stellar masses crudely estimated as 
described above.  As in Figure~3, we scale the virial products by the relevant 
$f$ factor (determined through the \msig\ relation) according to bulge type.  
Three things are immediately obvious, especially when we restrict our 
attention to the classical bulges and ellipticals that form an intrinsically 
well-defined and tight scaling relation among inactive galaxies. First, the 
AGN hosts lie systematically displaced from the inactive galaxies; at fixed 
$M_{\rm bulge}$, \mbh\ is $\sim$0.4 dex lower, or, equivalently, 
$M_{\rm bulge}$ is 0.35 dex higher at fixed \mbh.  Second, active hosts show 
greater intrinsic scatter ($0.52\pm0.09$ dex) than inactive systems 
($0.29\pm0.01$ dex; Kormendy \& Ho 2013).  And third, the largest offsets are 
confined preferentially to sources with the highest Eddington ratios 
(highlighted in blue). Dividing the sample by the median value of \lledd\ = 
0.07, objects above this limit have a median $\Delta \log M_{\rm bulge} 
\approx 0.5$ dex whereas those below this division have only 
$\Delta \log M_{\rm bulge} \approx 0.1$ dex.  

The offset between active and inactive galaxies in the \mmb\ plane lends 
itself to two possible explanations. One possibility is that active galaxies 
genuinely have systematically lower BH masses than inactive galaxies, or their 
BH masses have been systematically underestimated.  This was the interpretation
offered by Kim et al. (2008a) to account for a similar result seen in a sample 
of low-redshift quasars with available single-epoch BH masses and \hst\ 
photometry for the host.  This explanation can be ruled out for the current 
sample of RM AGNs, whose BH masses have been explicitly and independently 
calibrated through the \msig\ relation.  As discussed in Section~5.3, there 
might be a modest dependence of the $f$ factor on Eddington ratio, but this 
contributes only at the level of $\sim 0.2$ dex, not enough to account for the 
0.4 dex offset seen in Figure~5.  The second possibility is that the offset 
comes from a systematic overestimation of the stellar mass of the bulge due to 
the presence of young stars.  While the evidence for ongoing star formation in 
luminous AGNs remains controversial (e.g., Ho 2005; Kim et al. 2006; Shi et al.
2009; Mullaney et al. 2012), the association between nuclear activity and 
{\it recent}\ star formation is more secure (e.g., Cid~Fernandes et al. 2001; 
Kauffmann et al. 2003).  To explain the shift of $\sim$0.4 dex in 
$M_{\rm bulge}$ for the AGN population in Figure~5, we need to reduce the 
mass-to-light ratio by a factor of $2-2.5$ compared to the value we adopted for 
an evolved elliptical galaxy.  This implies that the integrated light of the 
stellar population of the bulges of AGN hosts resembles that of an Sbc or Sc 
spiral (Fukugita et al. 1995).  A qualitatively similar luminosity enhancement 
has been seen in other samples of local active galaxies (e.g., Nelson 
et al.  2004; Bennert et al. 2011; Busch et al. 2014).

\section{Summary and Future Directions}

Kormendy \& Ho (2013) recently show that the \msig\ and \mmb\ relations of 
nearby inactive galaxies depend strongly on bulge type.  Only classical bulges 
and elliptical galaxies obey these correlations tightly; pseudobulges show 
markedly different correlations characterized by a lower zero point and larger 
intrinsic scatter.  The virial mass estimators widely used to derive BH masses 
for broad-line AGNs require a zero point determination of the so-called $f$ 
factor that is established by scaling the virial product, VP = 
$r (\Delta V)^2/G$, derived from reverberating mapping, to the \msig\ relation 
of inactive galaxies.  Previous studies, while recognizing the possible 
importance of distinguishing between pseudo and classical bulges, have not 
explicitly attempted to take this into account.  

This paper remedies this situation by performing a comprehensive 
classification of the bulge type of the host galaxies of the latest sample of
reverberation-mapped AGNs, through careful assessment of all available 
high-resolution ground-based and space-based images and other ancillary data.
A total of 44 sources are included, of which 31 (16 pseudobulges and 15 
classical bulges and ellipticals) have reliable stellar velocity dispersion 
measurements.  We demonstrate, for the first time, that classical and pseudo 
bulges in active galaxies---as in inactive systems---obey distinctly different 
\msig\ relations.  At the same time, the slopes of the respective \msig\ 
relations for the two bulge types are, within the relatively large 
uncertainties, compatible between AGN and non-AGN hosts.  This justifies the 
use of the inactive galaxy sample to fix the zero point of the virial 
coefficient for RM AGNs.

Using the \msig\ relation of Kormendy \& Ho as the fiducial reference, we 
derive an updated $f$ factor appropriate for AGNs hosted by classical bulges 
and ellipticals.   We extend the calibration to pseudobulge AGN hosts using an 
approximate extrapolation of the \msig\ relation for inactive galaxies to 
pseudobulges systems.  In the case of virial products based on $\Delta V = 
\sigma_{\rm line}$(\hb) measured from rms spectra, $f=6.3\pm1.5$ for classical 
bulges and $f=3.2\pm0.7$ for pseudobulges.  While the $f$ factor for classical 
bulges is not substantially different from $f=5.5\pm1.7$ (Onken et al. 2004) 
or $5.2\pm1.2$ (Woo et al. 2010) commonly adopted in recent years, the value 
for pseudobulges is notably lower.

Our analysis shows that the $f$ factor, and, by implication, the structure and 
kinematics of the BLR, correlates with the large-scale properties of the host 
galaxy bulge.  While the two bulge types signify different evolutionary 
pathways (Kormendy \& Kennicutt 2004; Kormendy \& Ho 2013)---rapid formation 
via major mergers for classical bulges versus slow, secular evolution for 
pseudobulges---it is puzzling that the global conditions of the host connect 
to the physical properties of the nucleus on ultra-small scales.   Within each 
bulge type there is also tentative evidence that the BLR structure changes 
systematically with Eddington ratio.

Building on earlier work, this paper follows the conventional strategy of 
calibrating the average $f$ factor by scaling the VP of RM AGNs so that 
statistically they match the \msig\ relation of local inactive galaxies.  
However, in spite of a decade of observational effort (e.g., Ferrarese et al. 
2001; Nelson et al. 2004; Onken et al. 2004; Dasyra et al. 2007; Watson et al. 
2008; Woo et al. 2010, 2013), the number of RM AGNs with measured $\sigma_*$ 
(31) remains frustratingly small, especially that now, as we advocate, the 
sample must be subdivided into two groups according to bulge type.  The 
problem is particularly acute at the upper end of the mass distribution, which 
is most relevant for the quasar regime.  For example, of the 12 
objects in the RM sample with BH masses formally \gax\ $10^{8.5}$ \solmass\ 
(adopting $f = 6.3$), only three (3C~390.3, PG~1617+175, and PG~1426+015) have 
$\sigma_*$ measured.  The region of the \msig\ relation for AGNs with 
$\sigma_*$ \gax\ 200 \kms\ is dangerously underpopulated (Figure~3).  We are 
not optimistic that the situation will improve any time soon, as the most 
recent efforts to address this subject (e.g., Grier et al. 2013b) have already 
pushed 8--10~m class telescopes to their practical limits.  Dynamical modeling 
of velocity-resolved RM data may eventually yield robust individual BH masses 
without the need to invoke $f$, but this method is still in its infancy 
(Brewer et al. 2011; Pancoast et al. 2012; Li et al. 2013).

Where do we go from here?  

We suggest that the \mmb\ relation offers a promising substitute for the 
\msig\ relation.  As discussed in Kormendy \& Ho and in Section~2, the 
intrinsic scatter of both scaling relations is essentially indistinguishable. 
Thus, either one can serve as an effective anchor to calibrate $f$.  The 
stellar light of AGN host galaxies is generally much easier to measure 
photometrically than spectroscopically.  This can be seen from a simple 
comparison of the number of objects in Figure~5 versus Figure~3.  However, a 
couple of caveats should be borne in mind.  First, although high-resolution 
imaging can routinely detect host galaxy emission in relatively nearby AGNs 
(e.g., Kim et al. 2008a; Bentz et al. 2009a, 2013; Bennert et al. 2010), a 
variety of systematic uncertainties often makes quantitative measurements 
tricky (see Appendix~A), especially as they pertain to the bulge component 
that is maximally affected by the central point source (Kim et al. 2008b).  The 
S\'ersic index, a key ingredient to classify the bulge type, is particularly 
hard to determine accurately.  Second, we measure not stellar mass but light.  
Figure~5 reminds us that AGN hosts often contain nonnegligible quantities 
of young stars, and these must be properly quantified with multi-band colors 
and stellar population models before the \mmb\ relation can be used as a 
tool to constrain $f$.

\acknowledgements
We are grateful to an anonymous referee and Yue Shen for helpful comments.  We 
thank Misty Bentz for correspondence regarding the image decomposition 
presented in Bentz et al. (2013) and Daeseong Park for discussions about the 
linear regression fitting method.  This work is a direct outgrowth of the 
ARA\&A review LCH wrote with John Kormendy; LCH is deeply indebted to Kormendy 
for their fruitful collaboration.  This research was supported by NASA grants 
HST-AR-12818 and HST-GO-12903 from the Space Telescope Science Institute 
(operated by AURA, Inc., under NASA contract NAS5-26555).  LCH acknowledges 
additional support from the Kavli Foundation, Peking University, and the 
Chinese Academy of Science through grant No. XDB09030102 (Emergence of 
Cosmological Structures) from the Strategic Priority Research Program.  MK was 
partly supported by the KASI-Carnegie Fellowship Program, which is jointly 
managed by Korea Astronomy and Space Science Institute and the Observatories 
of the Carnegie Institution for Science.  We made extensive use of the 
NASA/IPAC Extragalactic Database (NED) which is operated by the Jet Propulsion 
Laboratory, California Institute of Technology, under contract with NASA.  


\clearpage
\appendix

\section{Notes on Individual Objects}

This section gives comments on the morphological characteristics of individual 
sources that are pertinent to classification of their bulge type.

\bigskip

{\it  3C 120       } --- \ \  This Fanaroff-Riley (1974) class I radio galaxy 
(Owen \& Laing 1989) is classified as an S0 in the Third Reference Catalogue 
of Bright Galaxies (RC3; de Vaucouleurs et al. 1991) and in Slavcheva-Mihova 
\& Mihov (2011), based on ground-based CCD images.  Analysis of an \hst\ 
ACS/HRC F550M image by Bentz et al. (2009a) shows that the bulge has a modest 
luminosity fraction of only $B/T = 0.21$ and a low \ser\ index of $n = 1.1$.  
Unfortunately, neither result is unambiguous.  An independent study by K14, 
using an ACS/HRC F702W image, yields a consistent value of $B/T = 0.19$, but a 
single-component model (i.e. $B/T = 1$) cannot be excluded; the \ser\ index 
could not be well constrained and was fixed to $n = 4$.  Bennert et al. (2010) 
also conclude that $B/T = 1$.   3C~120 is either an elliptical or 
a classical bulge, and, for concreteness, we choose the former.  

{\it  3C 390.3     } --- \ \ As in Bennert et al. (2010), K14 describe this 
Fanaroff-Riley class II radio galaxy (Laing et al. 1983) as an elliptical, 
consistent with $n = 4$.  Bentz et al. (2009a) conclude that a disk component 
is present ($B/T = 0.48$), although they also find a large \ser\ index of $n 
\approx 4$.  For concreteness, we assign a classification of elliptical.

{\it  Ark 120      } --- \ \ Ground-based morphological classifications of 
Ark~120 are uncertain, ranging from ``S?'' in RC3, ``Sb pec'' in Chatzichristou 
(2000), and E in Hyperleda.  Xanthopoulous (1996) finds $B/T = 0.89$ 
from decomposition of an $I$-band image.  All extant studies based on \hst\ 
images agree that the galaxy contains a substantial disk component, with 
$B/T$ ranging from $0.25\pm0.14$ (K14), to 0.35 (Bennert et al. 2010), 
to 0.49 (Bentz et al. 2009a).  The relatively large $B/T$, coupled with 
$n \approx 4$ (Bentz et al. 2009a; K14) suggests that Ark~120 has a classical 
bulge.

{\it  Arp 151      } --- \ \ Any formal classification of this object is made 
uncertain by the fact that the galaxy is stretched into a long tidal feature 
(Bentz et al. 2013; K14).  Although the bulge light accounts for only 
$25\pm15$\% of the total light, the bulge component formally has a fairly 
large \ser\ index of $n = 4.5\pm2.5$ (K14).  We classify the bulge as 
classical.

{\it  Fairall 9    } --- \ \ Kotilainen et al. (1993) obtain $B/T = 0.83$ from 
$R$-band imaging with 1\farcs27 seeing.  Using \hst\ photometry, Bentz et al. 
(2009a) and K14 also find the galaxy to be bulge-dominated, with $B/T = 0.52$ 
and $B/T = 0.64\pm0.29$, respectively; both find a large \ser\ index, 
$n \approx 4-6$.  We classify the bulge as classical.

{\it  Mrk 50       } --- \ \ All ground-based studies concur that Mrk~50 is 
bulge-dominated (McKenty 1990; Koss et al. 2011), in agreement with the visual 
appearance of its Sloan Digital Sky Survey (SDSS; York et al. 2000) image. K14 
obtain a good fit with $B/T = 0.89^{+0.11}_{-0.76}$ and $n = 2.0\pm1.0$.  We 
classify the bulge as classical.

{\it  Mrk 79       } --- \ \ A bar is prominently visible in large-scale, 
ground-based optical images (MacKenty 1990; Schmitt \& Kinney 2000; SDSS), and 
RC3 classifies the galaxy as an SBb.  Bennert et al. (2010) derive $B/T = 0.41$
from decomposition of an \hst\ image; somewhat lower values are found by Bentz 
et al. (2009a; $B/T = 0.19$) and K14 ($B/T = 0.16\pm0.09$), both of whom agree 
that $n \approx 3$.  In view of the relatively large value of $n$, we classify 
the bulge as classical.

{\it  Mrk 202      } --- \ \ Although suspected to be an elliptical in 
HyperLeda, the SDSS image suggests that Mrk~202 is more likely a nearly face-on 
early-type disk galaxy.  No obvious large-scale spiral arms are visible in the 
WFC3 F547M image of Bentz et al. (2013); instead, the central region contains 
a prominent circumnuclear ring.  Excluding the circumnuclear ring, the bulge 
accounts for 35\% of the total light of the host.  The \ser\ index of the bulge
is model-dependent and somewhat uncertain, lying in the range $n \approx 1-2$.  
We tentatively classify Mrk~202 as hosting a pseudobulge.

{\it  Mrk 509      } --- \ \ Motivated by the detection of a 4--5 kpc diameter 
ring-like feature, K14 fit the PC F547M image of Mrk~509 using a two-component, 
bulge plus disk model. The best fit formally yields $B/T = 0.16\pm0.08$, but 
it is highly unreliable because the bulge is barely resolved ($R_e$ = 
0\farcs13$\pm$0\farcs10), very faint relative to the nucleus (by $\sim 2.4$ 
mag), and suspiciously low in luminosity ($L_R = 10^{9.83}$ \solum) for its BH 
mass of $1.4\times 10^8$ \solmass.  The \ser\ index could not be independently 
constrained and was fixed to $n = 4$.  By contrast, Bentz et al. (2009a) 
conclude that $B/T = 1.0$, although their best fit indicates that $n = 1$.  
Evidence for an E-like morphology also comes from the ground-based work of 
MacKenty (1990) and Fischer et al. (2006).  We agree with Bentz et al. and 
classify Mrk~509 as an elliptical.

{\it  Mrk 590      } --- \ \ Images at ground-based resolution indicate that 
Mrk~590 is a bulge-dominated, early-type (Sa) spiral (MacKenty 1990; 
Kotilainen et al.  1993; Xanthopoulous 1996; SDSS).  \hst-based decompositions
yield more modest bulge fractions, from $B/T = 0.17-0.28$ ($n = 0.59$; Bentz 
et al. 2009a) to $B/T = 0.34$ ($n$ fixed to 4; Bennert et al. 2010).  K14 
quote $B/T = 0.26\pm0.11$ and $n = 1.04\pm0.52$. The near-exponential profile 
of the bulge formally designates it as a pseudobulge.

{\it  Mrk 1310     } --- \ \ With the \ser\ index fixed to $n = 4$, Schade 
et al. (2000) derive  $B/T = 0.23$ and, using the same WFPC2 F814W image,  K14 
find $B/T = 0.11\pm0.06$.  Very faint spiral arms can be seen in the residual 
image of K14.  Bentz et al. (2013) analyze a higher-resolution ACS/HRC F547M 
image and find $n = 4.8$ and $B/T = 0.46$.  In light of the large \ser\ index 
and relatively large $B/T$, we classify the bulge of Mrk~1310 as classical.

{\it  NGC 3227     } --- \ \ Classified as SAB(s)a in the RC3, Kormendy \& Ho 
regard NGC~3227 as having a pseudobulge because of the evidence of 
circumnuclear star formation (Davies et al. 2006).  They obtain a low value of 
$B/T = 0.108$ in the $K$ band, consistent with the value of $B/T = 0.098$ 
based on the $R$-band decomposition of Virani et al. (2000).  Interestingly, 
the bulge is significantly more prominent in \hst-based decompositions.  Bentz 
et al. (2009a) use an ACS/HRC F550M image to find $B/T = 0.65-0.74$ and 
$n = 2.14$, while K14's analysis of an ACS/HRC F606W image finds $B/T = 
0.34\pm0.18$ and $n = 3.19\pm1.60$.  The large discrepancy between the \hst\ 
and ground-based results most likely stems from the small field-of-view (FOV = 
29\asec$\times$26\asec) of the ACS/HRC images, which only cover the central 
region of this large galaxy ($D_{25}$ = 3\farcm9).  We tentatively follow 
Kormendy \& Ho and classify the bulge as a pseudobulge, and we adopt their 
ground-based $K$-band bulge luminosity and convert it to the $R$ band 
assuming $R-K$ = 2.37 [from $V-K = 2.98$ (Kormendy \& Ho 2013) and 
$V-R = 0.61$ (Fukugita et al. 1995)].

{\it  NGC 3516     } --- \ \ Classified as (R)SB(s)0 in the RC3, NGC~3516 
has $B/T = 0.81$ according to the $R$-band ground-based decomposition of 
Virani et al. (2000).  Bentz et al. (2009a) obtain $B/T = 0.52-0.86$ and 
$n = 0.96$ from analysis of an ACS/HRC F550M image and ground-based images; 
K14's analysis of a WFPC2 F814W image finds $B/T = 0.31\pm0.18$ and $n = 
1.15\pm0.57$.  The near-exponential profile of the bulge formally designates 
it as a pseudobulge.

{\it  NGC 3783     } --- \ \ As with NGC~3227, the \hst\ decomposition of 
NGC~3783 [RC3 type (R$^\prime$)SB(r)ab; $D_{25}$ = 2\farcm1] is highly 
unreliable because of the small FOV of the ACS/HRC.  Bentz et al. 
(2009a) and K14 analyze the same F550M image.  While both studies agree that 
$n$ is low ($1.1-1.4$), the bulge luminosities are highly discrepant: Bentz et 
al. (2009a) get $B/T = 0.07-0.12$, while K14 quote $B/T = 0.46\pm0.25$.  The 
low \ser\ index formally designates it as a pseudobulge.

{\it  NGC 4051     } --- \ \ As with NGC~3227, the \hst\ decomposition of 
NGC~4051 [RC3 type SAB(rs)bc; $D_{25}$ = 4\farcm9] is also highly unreliable 
because of the small ACS/HRC field.  The decompositions of Bentz et al. (2009a)
and K14 are in relatively good agreement, both finding $B/T \approx 0.4$ and 
$n \approx 1-2$.  The ground-based decomposition of Virani et al. (2010) 
requires a bulge only 25\% as luminous, suggesting that the \hst-based bulge 
luminosity may be biased too high as a result of the small FOV of the HRC.  
[Note, however, that Bentz et al. (2009a) did incorporate a larger FOV 
ground-based image into their fit.]  If the \ser\ index can be trusted, its 
low value formally designates NGC~4051 as having a pseudobulge.

{\it  NGC 4151     } --- \ \  The main bulge component of NGC~4151 accounts 
for 60\% of the total host galaxy light in the ACS/HRC F550M filter, but the 
formal \ser\ index is only $n = 0.81$ (Bentz et al. 2009a).  Bentz et al. 
introduce an additional inner bulge component that seems to make up another
15\% of the light; their fit includes large-FOV ground-based data.  K14 look 
at the same HRC data and conclude that the \ser\ index could not be fit 
reliably independently; fixing $n=4$, they derive $B/T = 0.32\pm0.17$.  
With $D_{25}$ = 2\farcm9, it is likely that this (R$^\prime$)SAB(rs)ab galaxy 
cannot be reliably fit using the small-FOV HRC image alone.  The ground-based 
$R$-band decomposition of Virani et al. (2000) gives $B/T = 0.44$.  Despite 
these uncertainties, it seems clear that NGC~4151 is a fairly bulge-dominated 
system, and thus that it probably has a classical bulge.

{\it  NGC 4253     } --- \ \ K14 obtain a reliable decomposition of NGC~4253 
(Mrk~766) using the full mosaic of WFPC2 in F606W: $B/T = 0.07\pm0.09$
and $n = 2.03\pm1.01$. These parameters are in good agreement with the results 
of Orban~de~Xivry et al. (2011), who, using the same dataset, find $B/T = 
0.11$ and $n = 1.88$.  They are also consistent with the analysis of Bentz 
et al. (2013) based on WFC3 F547M: $B/T \approx 0.05$ and $n = 1.1$ (their 
``simple'' model).  In view of the low $B/T$ and $n$, we classify the bulge of 
NGC~4253 as a pseudobulge.

{\it  NGC 4593     } --- \ \ Bentz et al. (2009a) derive $B/T = 0.71-0.76$ and 
$n = 1.94$ from modeling an HRC F550M image in combination with a larger FOV 
ground-based image.  The ground-based image is crucial because Mrk~1330 is 
large ($D_{25}$ = 2\farcm4). The same HRC dataset, analyzed by K14 without the 
aid of ground-based imaging, yields vastly different parameters for the bulge 
($B/T = 0.17$; $n = 0.17$.  We adopt the results of Bentz et al. (2009a), 
and given that the \ser\ index is formally, albeit marginally, less than 2, 
we classify the bulge as a pseudobulge.

{\it  NGC 4748     } --- \ \ The decomposition of this interacting early-type 
spiral is very complicated because it contains both a bar and a star-forming 
nuclear ring (Deo et al. 2006; Bentz et al. 2013).  Orban~de~Xivry et al. 
(2011) analyze a WFPC2 F606W image and find $B/T = 0.20$ and $n$ = 1.93.  
However, the more complex model of Bentz et al. (2013), based on WFC3 F547M 
data, gives $B/T \approx 0.1$ and $n = 4.8$ (their ``optimal'' model).  In 
spite of the large \ser\ index suggested by Bentz et al.'s analysis, we favor 
a pseudobulge classification for NGC~4748 because of its low $B/T$ and the 
presence of a nuclear ring.

{\it  NGC 5548     } --- \ \ The ground-based decomposition of Virani et al. 
(2000) yields $B/T = 0.57$ in the $R$ band.  This is consistent with 
the work of Bentz et al. (2009a), who, combining ground-based and HRC F550M 
data, report $B/T = 0.60-0.87$ and $n = 1.39$.  K14 analyze the same HRC 
dataset but find $B/T = 0.39\pm0.17$ and $n = 4.13\pm2.07$; the parameters of 
this component agree quite well with those of the ``inner bulge'' component 
in Bentz et al. (2009a). However, the latter authors detect an additional, 
larger component with 2.3 times the flux that they also attribute to the 
``bulge''.  In spite of the uncertainty on $n$, the large $B/T$ strongly 
suggests that NGC~5548 contains a classical bulge.  We adopt $B/T = 0.87$, the 
sum of the ``bulge'' and ``inner bulge'' components given in Bentz et al. 
(2009a).

{\it  NGC 6814     } --- \ \ The ground-based decomposition of Virani et al. 
(2000) yields $B/T = 0.26$ in the $R$ band.  Bentz et al. (2013) model a 
WFC3 F550M image with two bulge components, a bar, and a disk.  The two 
bulge components were fit with a \ser\ index of $n = 1.5$ and 1.2, 
respectively, and they both account for $\sim 8$\% of the total host 
galaxy light.  In view of the low values of $B/T$ and $n$, we classify 
NGC~6814 as having a pseudobulge.

{\it  NGC 7469     } --- \ \ This (R$^\prime$)SAB(rs)a galaxy contains a 
well-known starburst circumnuclear ring (e.g., Wilson et al. 1991; Deo et al. 
2006).  This characteristic, in conjunction with the fairly low \ser\ index 
of $n = 1.31$ found by Bentz et al. (2009a), suggests that the galaxy contains
a pseudobulge, even though the bulge is quite prominent ($B/T\approx 0.6-0.8$).

{\it  PG 0003+199  } --- \ \ All extant \hst-based decompositions find that 
PG~0003+199 (Mrk~335) is best fit with a bulge and disk with $B/T \approx 
0.2-0.4$ (Bentz et al. 2009a; Bennert et al. 2010), although K14 note that a 
single-component fit is also acceptable.  The host galaxy, whatever its form, 
is quite featureless (Crenshaw et al. 2003; K14).   The bulge component 
appears to have a fairly large, if poorly determined, \ser\ index.  We adopt 
K14's parameters, $B/T = 0.23\pm0.13$ and $n = 2.85\pm2.28$, and we classify 
the bulge as a classical bulge.

{\it  PG 0026+129  } --- \ \ All high-resolution studies agree that 
PG~0026+129 is hosted by an elliptical galaxy (McLeod \& McLeod 2001; Guyon 
et al. 2006; Bentz et al. 2009a; Veilleux et al. 2009; K14).  Bentz et al. 
(2009a) find a best-fitting \ser\ index of $n = 1.72$, but K14 could not 
reliably constrain it and fix it to $n = 4$.  In any case, the status of the 
host as an elliptical is not in doubt.

{\it  PG 0052+251  } --- \ \ This quasar has nearby neighbors and 
evident spiral structure (Dunlop et al. 2003; K14).  Single-component 
models for the host galaxy (e.g., Bennert et al. 2010) are thus disfavored.  
Bentz et al. (2009a) obtain $B/T = 0.32$ and $n = 3.12$ (HRC F550M). K14 
conclude that the \ser\ index cannot be well constrained; fixing $n = 4$, 
they find $B/T = 0.67^{+0.33}_{-0.38}$ (WF2 chips, F675W).  This is in 
excellent agreement with $B/T = 0.7$ derived by Dunlop et al. (2003) in the 
$R$ band.  We classify the bulge as classical.

{\it  PG 0804+761  } --- \ \ Guyon et al. (2006) observed this source with 
ground-based near-infrared imaging assisted by adaptive optics.  They 
find the host to be strongly bulge-dominated: $B/T = 0.82$ and 0.68 in the $H$ 
and $K$ band, respectively.  The three \hst\ studies of Bentz et al. 
(2009a), Bennert et al. (2010), and K14 all agree that the host is an 
elliptical ($B/T = 1$), although curiously the host seems to have an 
exponential profile ($n \approx 1$).

{\it  PG 0844+349  } --- \ \  As with PG~0052+251, PG~0844+349 contains 
very complex spiral structure (Hutchings \& Crampton 1990; Guyon et al. 2006; 
Veilleux et al. 2009), which is visible even in SDSS.  Thus, an elliptical 
galaxy model (Bennert et al. 2010) is inappropriate.  Bentz et al. (2009a) use 
an HRC F550M image to derive $B/T = 0.45$ and $n = 2.28$; K14 analyze the 
same data and find, fixing $n$ to 4, $B/T = 0.64\pm0.36$, consistent with 
$B/T = 0.69$ obtained from adaptive optics-based $H$-band observations (Guyon 
et al. 2006).  We classify the bulge as classical.

{\it  PG 0921+525  } --- \ \ Both Bentz et al. (2009a) and Bennert et al. 
(2010) fit the host with a two-component model, finding that $B/T = 0.17$ 
and 0.31, respectively.  The relatively low bulge fraction and low \ser\ index 
($n = 1.35$) led Bentz et al. to conclude that the host has a morphological
type Sc.  However, no spiral structure is visible.  While a two-component 
model is acceptable, K14 obtained a good fit of the host with a single 
component with $n$ fixed to 4.  We tentatively conclude that PG 0921+525 is 
an elliptical.

{\it  PG 0953+414  } --- \ \ All recent high-resolution studies agree that 
the host is an elliptical (Dunlop et al. 2003; Kim et al. 2008a; Bentz et al. 
2009a), or at least a strongly bulge-dominated (Guyon et al. 2006), galaxy.  
Bentz et al. (2009a) find a surprisingly low \ser\ index of $n = 1.39$.  K14 
conclude from their decomposition that the \ser\ index is difficult to 
constrain robustly.  The residual image from K14 shows evidence for a 
shell-like structure reminiscent of a tidal feature.

{\it  PG 1022+519  } --- \ \ Also known as Mrk~142, this system has a very 
strong bar that clearly shows up even in ground-based images (Ohta et al. 
2007; SDSS).  Its bulge, if present at all, is undetectable in the WFC3 F550W 
image of Bentz et al. (2013).  These authors further note that this object has 
a poorly determined RM lag.  We thus exclude it from our final analysis.

{\it  PG 1211+143  } --- \ \ Both Bentz et al. (2009a) and Bennert et al. 
(2010) analyze an ACS/HRC F550M image and conclude that $B/T = 1$, with the 
former further noting that $n = 1$.  K14 study an ACS/WFC F625W image and 
find a consistently large bulge ($B/T = 0.72^{+0.28}_{-0.41}$) but a 
significantly larger \ser\ index ($n = 5.84\pm4.67$).  We classify the bulge 
as classical.

{\it  PG 1226+023  } --- \ \ This well-studied quasar, 3C~273, is without a 
doubt hosted by an elliptical (Bentz et al. 2009a; Bennert et al. 2010; K14).

{\it  PG 1229+204  } --- \ \ The host appears to be barred (Surace et al. 
2001) and may contain a faint tidal tail (K14) and other signs of asymmetries 
(Veilleux et al. 2009).  The galaxy has a modest bulge, with $B/T \approx 
0.2-0.4$ and $n \approx 1.2-1.5$ (Bentz et al. 2009a; Bennert et al. 2010; 
K14).  We classify the bulge as a pseudobulge.

{\it  PG 1307+085  } --- \ \  The host is symmetric and featureless 
(Veilleux et al. 2009) and well fit with a single-component E-like model 
(Bentz et al. 2009a; K14).

{\it  PG 1351+695  } --- \ \ The galaxy, also known as Mrk~279, has a modest 
bulge with $B/T \approx 0.2-0.4$ and $n \approx 1.3-2.2$ (Bentz et al. 2009a; 
Bennert et al. 2010; Orban~de~Xivry et al. 2011; K14).  Faint spiral features 
are visible in the residual map of K14. We classify the bulge as a pseudobulge.

{\it  PG 1411+442  } --- \ \ Both Bentz et al. (2009a) and Bennert et al. 
(2010) fit the host with a single-component, E-like model.  This model, 
however, is inappropriate, as PG~1411+442 is clearly a highly distorted spiral 
galaxy (Guyon et al. 2006; Veilleux et al. 2009) with an extended tidal loop 
(Surace et al. 2001) and nearby companions (K14). K14's analysis of an ACS/WFC
F606W image finds $B/T = 0.42\pm0.23$; the \ser\ index could not be well 
constrained and was fixed to $n= 4$.  Guyon et al. (2006) find $B/T = 0.61$ and 
0.96 for the $H$ and $K$ band, respectively.  In light of the relatively 
large $B/T$, we classify the bulge as classical.

{\it  PG 1426+015  } --- \ \ The overall morphology of the host is strongly 
disturbed and asymmetric (Guyon et al. 2006; Veilleux et al. 2009; SDSS), 
rendering any detailed decomposition somewhat difficult to interpret.  While
the global profile is consistent with a single, E-like component (Guyon et al. 
2006; Bentz et al. 2009a), a bulge+disk decomposition is also acceptable.  
Assuming $n = 4$, Schade et al. (2000) obtain $B/T = 0.54$ in F814W, while 
K14, analyzing the same data, find $B/T = 0.30\pm0.14$ and $n = 2.12\pm1.70$.  
We believe that the ``disk'' component is actually tidally distorted 
material drawn from the main body of an elliptical host.

{\it  PG 1434+590  } --- \ \ The host, also known as Mrk~817, shows a clear 
bar and spiral arms (Crenshaw et al. 2003; Deo et al. 2006).  The bulge is 
tiny, comprising only $\sim 5$\% to 10\% of the total light of the host 
(Bentz et al. 2009a; Bennert et al. 2010; K14).  Bentz et al. (2009a) quote a 
\ser\ index of $n = 2.44$, but K14 find a much smaller value of $n \approx 
0.2$.  Despite this inherent uncertainty, it seems safe to conclude that the 
host contains a pseudobulge.

{\it  PG 1534+580  } --- \ \ The host galaxy of Mrk~290 is quite featureless 
at \hst\ resolution (Deo et al. 2006), but, with the sole exception of 
Bennert et al.  (2011), most studies conclude that the best-fitting model for 
the host requires a bulge and a disk component.  Published fits include those 
of Orban~de~Xivry et al. (2011; $B/T = 0.47$, $n = 4.06$), Bentz et al. (2013; 
$B/T = 0.33$, $n = 2.3$), and K14 ($B/T = 0.43^{+0.57}_{-0.31}$, $n = 4.04\pm
2.02$).  We adopt the parameters of K14 and classify the bulge as classical.

{\it  PG 1613+658  } --- \ \ The host galaxy of this infrared-excess quasar 
(Mrk~876) is a highly asymmetric, advanced merger remnant (McLeod \& McLeod 
2001; Surace et al. 2001; Guyon et al. 2006; Kim et al. 2008a) with two visible 
nuclei (K14).  Despite the significant residuals that remain, the azimuthally 
average global profile of the system is consistent with that of a giant 
elliptical galaxy (Guyon et al. 2006; Bentz et al. 2009a).  As in PG~1426+015, 
it seems likely that the tidal features will eventually relax and be 
incorporated into the main body of the remnant.  Thus, despite the fact that 
K14 formally find $B/T = 0.35$, we classify the host as an elliptical.

{\it  PG 1617+175  } --- \ \ The overall structure of the host is consistent 
with that of a relatively smooth elliptical (Guyon et al. 2006; Kim et al. 
2008a; Bentz et al. 2009a; Veilleux et al. 2009; K14).

{\it  PG 1700+518  } --- \ \ The overall structure of the host is consistent 
with that of a highly disturbed elliptical or at least a strongly 
bulge-dominated system (Guyon et al. 2006; Kim et al. 2008a; Bentz et al. 2009a;
Veilleux et al. 2009; Bennert et al. 2010; K14).  We classify the host 
as an elliptical or as having a classical bulge.

{\it  PG 2130+099  } --- \ \ This moderately inclined spiral has an outer ring 
(Slavcheva-Mihova \& Mihov 2011) and an asymmetric outer disk (K14).  
Published bulge-to-disk decompositions span $B/T \approx 0.1-0.4$ (Guyon et 
al. 2006; Kim et al. 2008a; Bentz et al. 2009a; Bennert et al. 2010; K14). The 
\ser\ index of the bulge appears to be very low, $n \approx 0.5-0.6$ (Bentz et 
al. 2009a; K14).   We adopt the parameters from K14 and classify the bulge as 
a pseudobulge.  Bentz et al. (2013) note that the RM lag for this source may 
be unreliable; we flag this object in our analysis.

{\it  SBS 1116+583A} --- \ \ Bentz et al. (2013) fit this barred spiral with a 
model consisting of a small bulge, bar, lens, and exponential disk.  In their 
optimal model, $B/T = 0.20$ and $n = 1.1$.  We classify the bulge as a 
pseudobulge.


\begin{thebibliography}{}

\bibitem[]{} 
Barth, A.~J., Pancoast, A., Thorman, S. J., et al. 2011, \apj, 743, L4

\bibitem[]{} 
Bell, E.~F., McIntosh, D. H., Katz, N., \& Weinberg, M. D. 2003, \apjs, 149, 289

\bibitem[]{} 
Bennert, V. N., Auger, M. W., Treu, T., Woo, J.-H., \& Malkan, M. A. 2011, \apj, 726, 59

\bibitem[]{} 
Bennert, V. N., Treu, T., Woo, J.-H., et al. 2010, \apj, 708, 1507

\bibitem[]{} 
Bentz, M. C., Denney, K. D., Cackett, E. M., et al. 2006, \apj, 651, 775

\bibitem[]{} 
Bentz, M. C., Denney, K. D., Grier, C. J., et al. 2013, \apj, 767, 149

\bibitem[]{}
Bentz, M. C., Peterson, B. M., Netzer, H., Pogge, R. W., \& Vestergaard, M. 2009a, \apj, 697, 160

\bibitem[]{}
Bentz, M. C.,  Walsh, J. L., Barth, A. J., et al. 2009b, \apj, 705, 199

\bibitem[]{} 
Blandford, R.~D., \& McKee, C.~F. 1982, \apj, 255, 419

\bibitem[]{} 
Boroson, T.~A. 2002, \apj, 565, 78

\bibitem[]{} 
Brewer, B. J., Treu, T., Pancoast, A., et al. 2011, \apj, 733, L33

\bibitem[]{} 
Busch, G., Zuther, J., Valencia-S., M., et al. 2014, \aa, 561, A140

\bibitem[]{} 
Calzetti, D., Kinney, A.~L., \& Storchi-Bergmann, T. 1994, \apj, 429, 582

\bibitem[]{} 
Chatzichristou, E.~T. 2000, \apjs, 131, 71

\bibitem[]{} 
Cid Fernandes, R., Heckman, T. M., Schmitt, H. R., et al. 2001, \apj, 558, 81

\bibitem[]{} 
Collin, S., Kawaguchi, T., Peterson, B., \& Vestergaard, M. 2006, \aa, 456, 75

\bibitem[]{} 
Crenshaw, D.~M., Kraemer, S.~B., \& Gabel, J. R. 2003, \aj, 126, 1690

\bibitem[]{} 
Dasyra, K. M., Tacconi, L. J., Davies, R. I., et al. 2007, \apj, 657, 102

\bibitem[]{} 
Davies, R. I., Thomas, J., Genzel, R., et al. 2006, \apj, 646, 754

\bibitem[]{} 
Denney, K. D., Bentz, M. C., Peterson, B. M., et al. 2006, \apj, 653, 152

\bibitem[]{} 
Denney, K. D., Peterson, B. M., Pogge, R. W., et al. 2010, \apj, 721, 715

\bibitem[]{} 
Deo, R. P., Crenshaw, D. M., \& Kraemer, S. B. 2006, \aj, 132, 321

\bibitem[]{} 
de Vaucouleurs, G., de Vaucouleurs, A., Corwin Jr., H.~G., et al. 1991, Third Reference Catalogue of Bright Galaxies (New York: Springer) (RC3)

\bibitem[]{} 
Dibai, E. A. 1977, Sov. Astron. Lett., 3, 1

\bibitem[]{} 
Dibai, E. A. 1984, Sov. Astron., 28, 245

\bibitem[]{} 
Dietrich, M., Peterson, B. M., Grier, C. J., et al. 2012, \apj, 757, 53

\bibitem[]{} 
Fanaroff, B.~L., \& Riley, J.~M. 1974, \mnras, 167, 31P

\bibitem[]{} 
Ferrarese, L., \& Merritt, D. 2000, \apj, , 539, L9

\bibitem[]{} 
Ferrarese, L., Pogge, R.~W., Peterson, B.~M., et al. 2001, \apj, 555, L79

\bibitem[]{} 
Fischer, S., Iserlohe, C., Zuther, J., et al. 2006, \aa, 452, 827

\bibitem[]{} 
Fisher, D. B., \& Drory, N. 2008, \aj, 136, 773

\bibitem[]{} 
Fukugita, M., Shimasaku, K., \& Ichikawa, T. 1995, \pasp, 107, 945

\bibitem[]{} 
Gadotti, D.~A. 2009, \mnras, 393, 1531

\bibitem[]{} 
Gebhardt, K., Bender, R., Bower, G., et al. 2000a, \apj, 539, L13

\bibitem[]{} 
Gebhardt, K., Kormendy, J., Ho, L. C., et al. 2000b, \apj, 543, L5

\bibitem[]{} 
Graham, A. W., Onken, C. A., Athanassoula, E., \& Combes, F. 2011, \mnras, 412, 2211

\bibitem[]{} 
Greene, J. E., \& Ho, L. C. 2006, \apj, 641, L21

\bibitem[]{} 
Greene, J. E., Ho, L. C., \& Barth, A. J. 2008, \apj, 688, 159

\bibitem[]{} 
Grier, C. J., Martini, P., Watson, L. C., et al. 2013b, \apj, 773, 90

\bibitem[]{} 
Grier, C. J., Peterson, B. M., Horne, K., et al. 2013a, \apj, 764, 4

\bibitem[]{} 
Grier, C. J., Peterson, B. M., Pogge, R. W., et al. 2012, \apj, 755, 60

\bibitem[]{} 
G\"ultekin, K., Richstone, D. O., Gebhardt, K., et al. 2009, \apj, 698, 198

\bibitem[]{} 
Guyon, O., Sanders, D. B., \& Stockton, A. 2006, \apjs, 166, 89

\bibitem[]{} 
H\"aring, N., \&  Rix, H.-W. 2004, \apj, 604, L89

\bibitem[]{} 
Hicks, E.~K. S., \& Malkan, M.~A. 2008, \apjs, 174, 31

\bibitem[]{} 
Ho, L.~C. 1999, in Observational Evidence for Black Holes in the Universe, ed. 
S.~K. Chakrabarti (Dordrecht: Kluwer), 157

\bibitem[]{} 
Ho, L.~C. 2005, \apj, 629, 680

\bibitem[]{} 
Hutchings, J.~B., \& Crampton, D. 1990, \aj, 99, 37

\bibitem[]{} 
Jiang, Y.-F., Greene, J. E., Ho, L. C., Xiao, T., \& Barth, A. J. 2011, \apj, 742, 68

\bibitem[]{} 
Kaspi, S., Smith, P. S., Netzer, H., et al. 2000, \apj, 533, 631

\bibitem[]{} 
Kauffmann, G., Heckman, T. M., Tremonti, C., et al. 2003, \mnras, 346, 1055

\bibitem[]{} 
Kim, M., Ho, L. C., \& Im, M. 2006, \apj, 642, 702

\bibitem[]{} 
Kim, M., Ho, L. C., Peng, C. Y., et al. 2008a, \apj, 687, 767

\bibitem[]{} 
Kim, M., Ho, L. C., Peng, C. Y., Barth, A. J., \& Im, M. 2008b, \apjs, 179, 28

\bibitem[]{} 
Kim, M., Ho, L. C., Peng, C. Y., Barth, A. J., \& Im, M. 2014, in preparation (K14)

\bibitem[]{} 
Kinney, A.~L., Calzetti, D., Bohlin, R.~S., et al. 1996, \apj, 467, 38

\bibitem[]{} 
Komatsu, E., Dunkley, J., Nolta, M. R., et al. 2009, \apjs, 180, 330

\bibitem[]{} 
Kormendy, J., Bender, R., \& Cornell, M. E. 2011, Nature, 469, 374

\bibitem[]{} 
Kormendy, J., \& Gebhardt, K. 2001, in The 20th Texas Symposium on Relativistic Astrophysics, ed. H. Martel \& J.~C. Wheeler (Melville: AIP), 363

\bibitem[]{} 
Kormendy, J., \& Ho, L. C. 2013, ARA\&A, 51, 551

\bibitem[]{} 
Kormendy, J., \& Kennicutt, R. C. 2004, ARA\&A, 42, 603

\bibitem[]{} 
Kormendy, J., \& Richstone, D. O. 1995, ARA\&A, 33, 581 

\bibitem[]{} 
Koss, M., Mushotzky, R., Veilleux, S., et al. 2011, \apj, 739, 57

\bibitem[]{} 
Kotilainen, J.~K., Ward, M.~J., \& Williger, G.~M. 1993, \mnras, 263, 655

\bibitem[]{} 
Krolik, J.~H. 2001, \apj, 551, 72

\bibitem[]{} 
Laing, R.~A., Riley, J.~M., \& Longair, M.~S. 1983, \mnras, 204, 151

\bibitem[]{} 
Li, Y.-R., Wang, J.-M., Ho, L. C., Du, P., \& Bai, J.-M. 2013, \apj, 779, 110

\bibitem[]{} 
MacKenty, J.~W. 1990, \apjs, 72, 231

\bibitem[]{} 
Magorrian, J., Tremaine, S., Richstone, D., et al. 1998, AJ, 115, 2285

\bibitem[]{} 
McConnell, N. J., \& Ma, C.-P. 2013, \apj, 764, 184

\bibitem[]{} 
McConnell, N. J., Ma, C.-P., Gebhardt, K., et al. 2011, Nature, 480, 215

\bibitem[]{} 
McLeod, K.~K., \& McLeod, B.~A. 2001, \apj, 546, 782

\bibitem[]{}
McLure, R.~J., \& Dunlop, J.~S. 2004, \mnras, 352, 1390

\bibitem[]{} 
Mullaney, J. R., Pannella, M., Daddi, E., et al. 2012, \mnras, 419, 95

\bibitem[]{} 
Negrete, C. A., Dultzin, D., Marziani, P., \& Sulentic, J. 2013, \apj, 771, 31

\bibitem[]{} 
Nelson, C. H., Green, R. F., Bower, G., Gebhardt, K., \& Weistrop, D. 2004, \apj, 615, 652

\bibitem[]{} 
Nelson, C.~H., \& Whittle, M. 1995, \apjs, 99, 67

\bibitem[]{} 
Ohta, K., Aoki, K., Kawaguchi, T., \& Kiuchi, G. 2007, \apjs, 169, 1

\bibitem[]{} 
Onken, C. A., Ferrarese, L., Merritt, D., et al. 2004, \apj, 615, 645

\bibitem[]{} 
Onken, C.~A., Valluri, M., Peterson, B.~M., et al. 2007, \apj, 670, 105

\bibitem[]{} 
Orban de Xivry, G., Davies, R., Schartmann, M., et al. 2011, \mnras, 417, 2721

\bibitem[]{} 
Osterbrock, D.~E., \& Pogge, R.~W. 1985, \apj, 297, 166

\bibitem[]{} 
Owen, F.~N., \& Laing, R.~A. 1989, \mnras, 238, 357

\bibitem[]{} 
Pancoast, A., Brewer, B. J., Treu, T., et al. 2012, \apj, 754, 49

\bibitem[]{} 
Park, D., Kelly, B. C., Woo, J.-H., \& Treu, T. 2012a, \apjs, 203, 6

\bibitem[]{} 
Park, D., Woo, J.-H., Treu, T., et al. 2012b, \apj, 747, 30

\bibitem[]{} 
Peterson, B. M. 1993, PASP, 105, 247

\bibitem[]{} 
Peterson, B.~M., Ferrarese, L., Gilbert, K. M., et al.  2004, \apj, 613, 682

\bibitem[]{} 
Press, W. H., Teukolsky, S. A., Vetterling, W. T., \& Flannery, B. P. 1992,
Numerical Recipes in C: The Art of Scientific Computing (2nd ed.;
Cambridge: Cambridge Univ. Press)

\bibitem[]{} 
Schade, D.~J., Boyle, B.~J., \& Letawsky, M. 2000, \mnras, 315, 498

\bibitem[]{} 
Schmitt, H.~R., \& Kinney, A.~L. 2000, \apjs, 128, 479

\bibitem[]{} 
Sergeev, S.~G., Pronik, V.~I., Sergeeva, E.~A., \& Malkov, Yu.~F. 1999, \aj, 118, 2658

\bibitem[]{} 
S\'ersic, J.~L. 1968, Atlas de Galaxias Australes (C\'ordoba: Obs. Astron., 
Univ. Nac. C\'ordoba)

\bibitem[]{} 
Shi, Y., Rieke, G. H., Ogle, P., Jiang, L., \& Diamond-Stanic, A. M. 2009, \apj, 703, 1107

\bibitem[]{} 
Simien, F., \& de Vaucouleurs, G. 1986, \apj, 302, 564

\bibitem[]{} 
Sincell, M.~W., \& Krolik, J.~H. 1998, \apj, 496, 737

\bibitem[]{} 
Slavcheva-Mihova, L., \& Mihov, B. 2011, \aa, 526, A43

\bibitem[]{} 
Sluse, D., Hutsem\'ekers, D., Courbin, F., Meylan, G., \& Wambsganss, J. 2012, \aa, 544, A62

\bibitem[]{} 
Sun, W.-H., \& Malkan, M.~A. 1989, \apj, 346, 68

\bibitem[]{} 
Surace, J.~A., Sanders, D.~B., \& Evans, A.~S. 2001, \aj, 122, 2791

\bibitem[]{} 
Tremaine, S., Gebhardt, K., Bender, R., et al. 2002, \apj, 574, 740

\bibitem[]{} 
Vanden Berk, D.~E., Richards, G. T., Bauer, A., et al. 2001, \aj, 122, 549

\bibitem[]{} 
Veilleux, S., Kim, D.-C., Rupke, D. S. N., et al. 2009, \apj, 701, 587

\bibitem[]{} 
Virani, S., De Robertis, M.~M., \& VanDalfsen, M.~L. 2000, \aj, 120, 1739

\bibitem[]{} 
Wandel, A., Peterson, B.~M., \& Malkan, M.~A. 1999, \apj, 526, 579

\bibitem[]{} 
Watson, L. C., Martini, P., Dasyra, K. M., et al. 2008, \apj, 682, L21

\bibitem[]{} 
Wilson, A.~S., Helfer, T.~T., Haniff, C.~A., \& Ward, M.~J. 1991, \apj, 381, 79

\bibitem[]{} 
Woo, J.-H., Schulze, A., Park, D., et al. 2013, \apj, 772, 49

\bibitem[]{} 
Woo, J.-H., Treu, T., Barth, A. J., et al. 2010, \apj, 716, 269

\bibitem[]{} 
Xanthopoulos, E. 1996, \mnras, 280, 6

\bibitem[]{} 
Xiao, T., Barth, A. J., Greene, J. E., et al. 2011, \apj, 739, 28

\bibitem[]{} 
York, D.~G., Adelman, J., Anderson Jr., J. E., et al. 2000, \aj, 120, 1579

\bibitem[]{} 
Zu, Y., Kochanek, C. S., \& Peterson, B. M. 2011, \apj, 735, 80

\end{thebibliography}
\end{document}